\documentclass[12pt]{article}
\usepackage{graphicx,amsmath,amssymb}
\usepackage{indentfirst}
\usepackage[usenames]{color}
\usepackage[colorlinks=true, urlcolor=navyblue, linkcolor=navyblue, citecolor=navyblue]{hyperref}
\usepackage{epstopdf}
\usepackage{enumerate}
\usepackage{appendix}
\usepackage{lineno}
\usepackage{setspace}

\definecolor{navyblue}{rgb}{0,0.08,0.45}
\definecolor{darkred}{rgb}{0.7,0.0,0.0}
\definecolor{darkgreen}{rgb}{0,0.6,0.2}

\newcommand{\beq}{\begin{equation}}
\newcommand{\enq}{\end{equation}}
\newcommand{\beqa}{\begin{eqnarray}}
\newcommand{\beqast}{\begin{eqnarray*}}
\newcommand{\enqa}{\end{eqnarray}}
\newcommand{\enqast}{\end{eqnarray*}}

\newcommand{\bec}{\begin{center}}
\newcommand{\enc}{\end{center}}
\newcommand{\beqo}{\begin{quote}}
\newcommand{\enqo}{\end{quote}}
\newcommand{\bem}{\begin{minipage}}
\newcommand{\enm}{\end{minipage}}

\begin{document}

\vspace{15pt}

\begin{center}
{\large Color Confinement,  Hadron Dynamics,  and Hadron Spectroscopy from Light-Front Holography and Superconformal Algebra}

\vspace{10pt}


\end{center}

\vspace{15pt}

\centerline{Stanley J. Brodsky}

\vspace{3pt}

\centerline {\it SLAC National Accelerator Laboratory, Stanford University}

\vspace{10pt}

\begin{abstract}
The QCD light-front Hamitonian equation $H_{LF}|\Psi> = M^2 |\Psi>$ derived from quantization at fixed LF time $\tau = t+z/c$ provides a causal, frame-independent, method for computing hadron spectroscopy as well as dynamical observables such as structure functions, transverse momentum distributions, and distribution amplitudes.   The QCD Lagrangian with zero quark mass has no explicit mass scale. 
de Alfaro, Fubini, and Furlan (dAFF) have made an important observation that a mass scale can appear in the equations of motion without affecting the conformal invariance of the action if one adds a term to the Hamiltonian proportional to the dilatation operator or the special conformal operator.   If one applies the dAFF procedure to the QCD light-front Hamiltonian, it leads  to a color confining potential $\kappa^4 \zeta^2$ for mesons, where $\zeta^2$ is the LF radial variable conjugate to  the $q \bar q$ invariant mass squared.  The same result, including spin terms, is obtained using  light-front holography  -- the duality between light-front dynamics and AdS$_5$ --  if one  
modifies  the AdS$_5$ action by the dilaton $e^{\kappa^2 z^2}$ in the fifth dimension $z$.   When one generalizes this procedure using superconformal algebra, the resulting light-front eigensolutions provide a unified Regge spectroscopy of meson, baryon, and tetraquarks, including remarkable supersymmetric relations between the masses of mesons and baryons and a universal Regge slope. The pion $q \bar q$ eigenstate has zero mass at $m_q=0.$  The superconformal relations also can be extended to heavy-light quark  mesons and baryons.
AdS/QCD also predicts the analytic  form of the nonperturbative running coupling  $\alpha_s(Q^2) \propto e^{-{Q^2\over 4 \kappa^2}}$,  in agreement with the effective charge  measured from measurements of the Bjorken sum rule.  The mass scale $\kappa$ underlying hadron masses  can be connected to the parameter   $\Lambda_{\overline {MS}}$ in the QCD running coupling by matching the nonperturbative dynamics to the perturbative QCD  regime. The result is an effective coupling $\alpha_s(Q^2)$  defined at all momenta.   One also obtains empirically viable predictions for spacelike and timelike hadronic form factors, structure functions, distribution amplitudes, and transverse momentum distributions. \end{abstract}


\newpage

\section{Introduction}

A  profound question in hadron physics is how the proton mass and other hadronic mass scales can be determined by QCD since there is  no explicit  parameter with mass dimensions in the QCD Lagrangian for vanishing quark mass.    This dilemma is compounded by the fact that  the chiral QCD Lagrangian has no knowledge of the conventions used for units of mass such as $MeV$.     Thus QCD with $m_q=0$ can in principle only predict {\bf ratios of masses } such as $m_\rho/m_p$ -- not their absolute values. Similarly,  given that color is confined, how does QCD set its range without a parameter with dimensions of length?  It is hard to see how this mass scale problem could be solved by `` dimensional transmutation", since the mass scale determined by  perturbative QCD such as $\Lambda_{\overline MS}$, is renormalization-scheme dependent, whereas hadron masses  are independent of the conventions chosen to regulate the UV divergences.

A remarkable principle,  first demonstrated by  de Alfaro, Fubini and Furlan  (dAFF)~\cite{deAlfaro:1976je} for conformal theory in $1+1$ quantum mechanics, is that a mass scale can appear in a Hamiltonian and its equations of motion without 
affecting  the conformal invariance of the action.  The essential step introduced by dAFF is to add to the conformal Hamiltonian terms proportional to the dilation operator $D$ and the special conformal operator $K$.  The unique result is the addition of a harmonic oscillator potential $V(x) = \kappa^4 x^2$ to the Hamlitonian,     The group algebra is maintained despite the fact that $D$ and $K$ have dimensions,  In fact, the new Hamitonian has ``extended dilatation invariance"  since the mass scale $\kappa$ can be rescaled arbitrarily. This implies that only ratios of the mass eigenvalues can be determined, not their absolute values.

De T\'eramond, Dosch, and I~\cite{Brodsky:2013ar}
have shown that a mass gap and color confinement appears when one extends the dAFF procedure to relativistic, causal, Poincar\'e invariant, light-front Hamiltonian theory for QCD. 
The resulting predictions for both hadronic spectroscopy and dynamics provide an elegant description of meson and baryon phenomenology, including Regge trajectories with universal slopes in the principal quantum number $n$ and the orbital angular momentum $L$.  In addition, the resulting quark-antiquark bound-state equation predicts a massless pion for zero quark mass.  In this contribution I will review recent advances in holographic QCD, extending an earlier review given in ref.~\cite{Brodsky:2016vig}.

Light-Front quantization is the natural formalism for relativistic quantum field theory.  Measurements of hadron structure, such as deep inelastic lepton-proton scattering, are made at fixed light-front time $\tau= t+z/c$, analogous to a flash photograph, not at a single ``instant time".  As shown by Dirac~\cite{Dirac:1949cp}, boosts are kinematical in the ``front form".  Thus all formulae using the front form are independent of the 
observer's motion~\cite{Brodsky:1997de}; i.e., they are  Poincar\'e invariant.  The eigenstates of the light-front Hamiltonian $ H_{LF} =  P^+ P^- -{\vec P}^2_\perp$ derived from the QCD Lagrangian encodes the entire the hadronic mass spectrum for both individual hadrons and  the multi-hadron continuum.   The eigenvalues of the LF Hamiltonian  are the squares of the hadron masses $M^2_H$: 
$H_{LF}|\Psi_H>  = M^2_H |\Psi_H>$~\cite{Brodsky:1997de}.   The evaluation of the Wilson line for gauge theories in the front form is discussed in ref.~\cite{Reinhardt:2016fjl}.

The eigenfunctions of the light-front Hamiltonian $ H_{LF} =  P^+ P^- -{\vec P}^2_\perp$ derived from the QCD Lagrangian correspond to the single hadron and  multi-hadronic continuum eigenstates.   The eigenvalues of the LF Hamiltonian  are the squares of the hadron masses $M^2_H$: 
$H_{LF}|\Psi_H>  = M^2_H |\Psi_H>$~\cite{Brodsky:1997de}.
Here  $P^- = i {d\over d\tau}$ is the LF time evolution operator, and $P^+=P^0+P^z$ and $\vec P_\perp$ are kinematical.  
The eigenfunctions of $H_{LF} $ provide hadronic LF Fock state wavefunctions (LFWFs):
 $ \psi^H_n(x_i, \vec k_{\perp i },\lambda_i)= <n| \Psi_H> $, the projection of the hadronic eigenstate on the free Fock basis. The constituents' physical momenta are 
$p^+_i = x_i P^+$, and  $\vec p_{\perp i } =  x_i  {\vec P}_\perp +  \vec k_{\perp i }$,  and the $\lambda_i$ label the  spin projections $S^z_i$. Remarkably one can reduce the LF Hamiltonian theory for $q \bar q$ mesons with $m_q=0$   to an effective  LF Schrodinger equation in  a single variable, the LF radial variable 
$\zeta^2 = b^2_\perp x(1-x)$

The LFWFs are Poincar\'e invariant: they are independent of $P^+$ and $P_\perp$ and are thus independent of the motion of the observer.  Since the LFWFs are independent of the hadron's momentum, there is no length contraction~\cite{Terrell:1959zz,Penrose:1959vz}. Structure functions are essentially the absolute square of the LFWFs.
One thus measures the  same structure function in an electron-ion collider as in an electron-scattering experiment where the target hadron is at rest.

\begin{figure}
 \begin{center}
\includegraphics[height= 12cm,width=15cm]{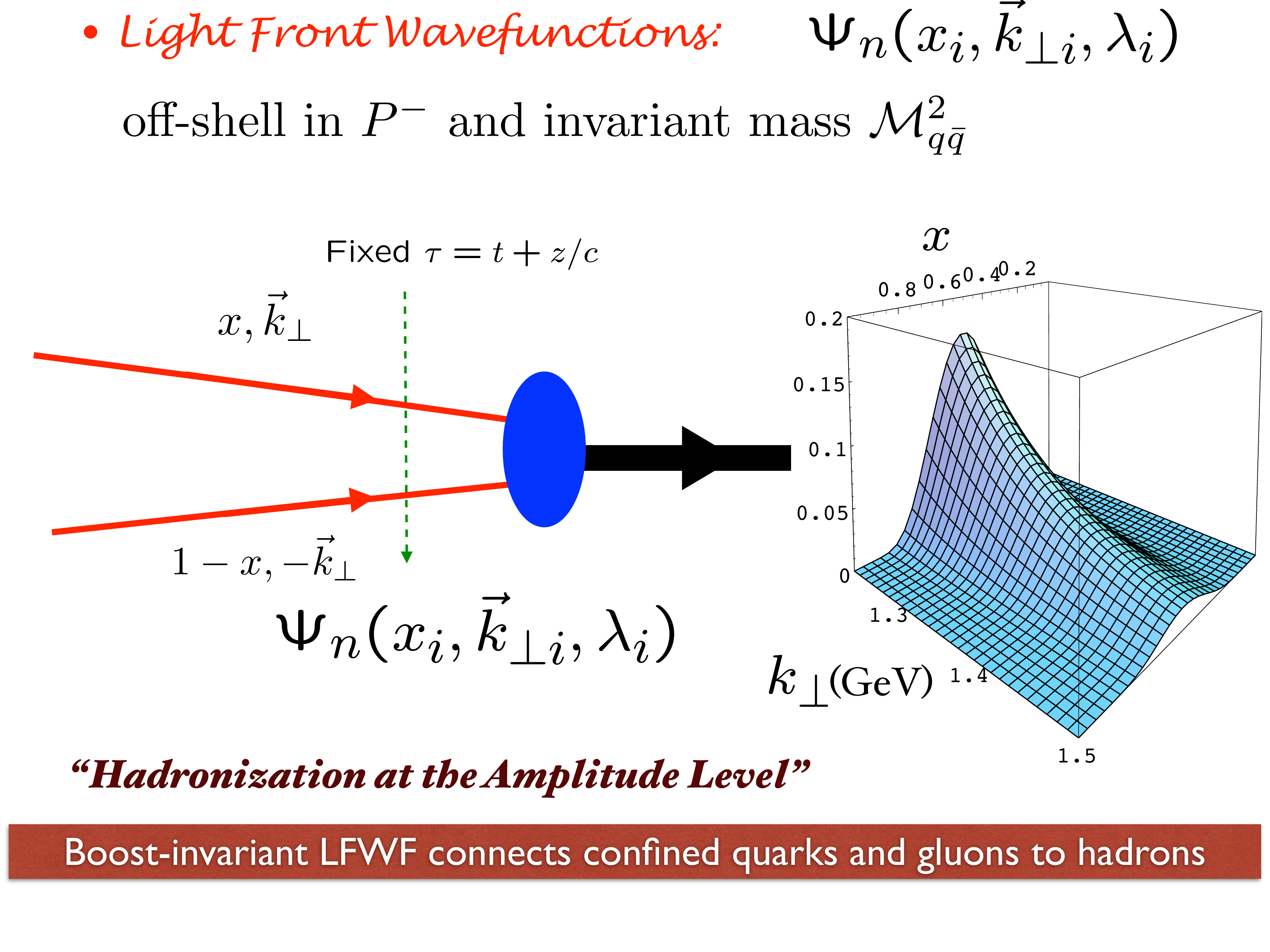}
\end{center}
\caption{ The meson LFWF connects the intermediate $q \bar q $ state, which is off of the $P^-$ energy shell and thus off-the-invariant mass shell  ${\cal M}^2  > m^2_H$T to the  physical meson state with 
${\cal M}^2  = m^2_H$.    The $q$ and $\bar q$ can be regarded as effective dressed fields}. 
\label{had1}
\end{figure}

Light-front wavefunctions thus provide a direct link between the QCD Lagrangian and hadron structure. Since they are defined at a fixed $\tau$, they connect the physical on-shell hadronic state to its quark and gluon parton constituents, not at off-shell energy, but off-shell in invariant mass squared ${\cal M}^2 =( \sum_i k^\mu_i )^2.$ 
They thus control the transformation of the quarks and gluons in an off-shell intermediate state into the observed final on-shell hadronic state. See fig. \ref{had1}.

One of the most elegant features of quantum field theory is supersymmetry --  where fermionic and bosonic eigensolutions  have the same mass. 
The conformal group has an elegant $ 2\times 2$ Pauli matrix representation called superconformal algebra, originally discovered by  Haag, Lopuszanski, and Sohnius.Ä ~\cite{Haag:1974qh}(1974)
The conformal Hamiltonian operator and the special conformal operators can be represented as anticommutators of Pauli matrices
$H = {1/2}[Q, Q^\dagger]$ and  $K = {1/2}[S, S^\dagger]$.
As shown by Fubini and Rabinovici,~\cite{Fubini:1984hf},  a nonconformal Hamiltonian with a mass scale and universal confinement can then be obtained by shifting $Q \to Q +\omega K$, the analog of the dAFF procedure. Thus the conformal algebra can be extended even though $\omega$ has dimension of mass.  
In effect one has generalized supercharges of the superconformal algebra~\cite{Fubini:1984hf}.   The result of this shift of the Hamiltonian is a color-confining harmonic  potential in the equations of motion.  Remarkably the action remains conformally invariant, and only one mass parameter appears.   

As shown by Guy de T\'eramond, G\"unter Dosch and myself, the bound-state equations of superconformal algebra are, in fact, Lorentz invariant, frame-independent, relativistic light-front Schrodinger equations, 
and the resulting eigensolutions are the eigenstates of a light-front Hamiltonian obtained from $AdS_5$  and light-front holography.  Light-front quantization at fixed light-front time $\tau=t+z/c$  provides a physical, frame-independent formalism for hadron dynamics and structure.   

\begin{figure}
 \begin{center}
\includegraphics[height=12cm,width=15cm]{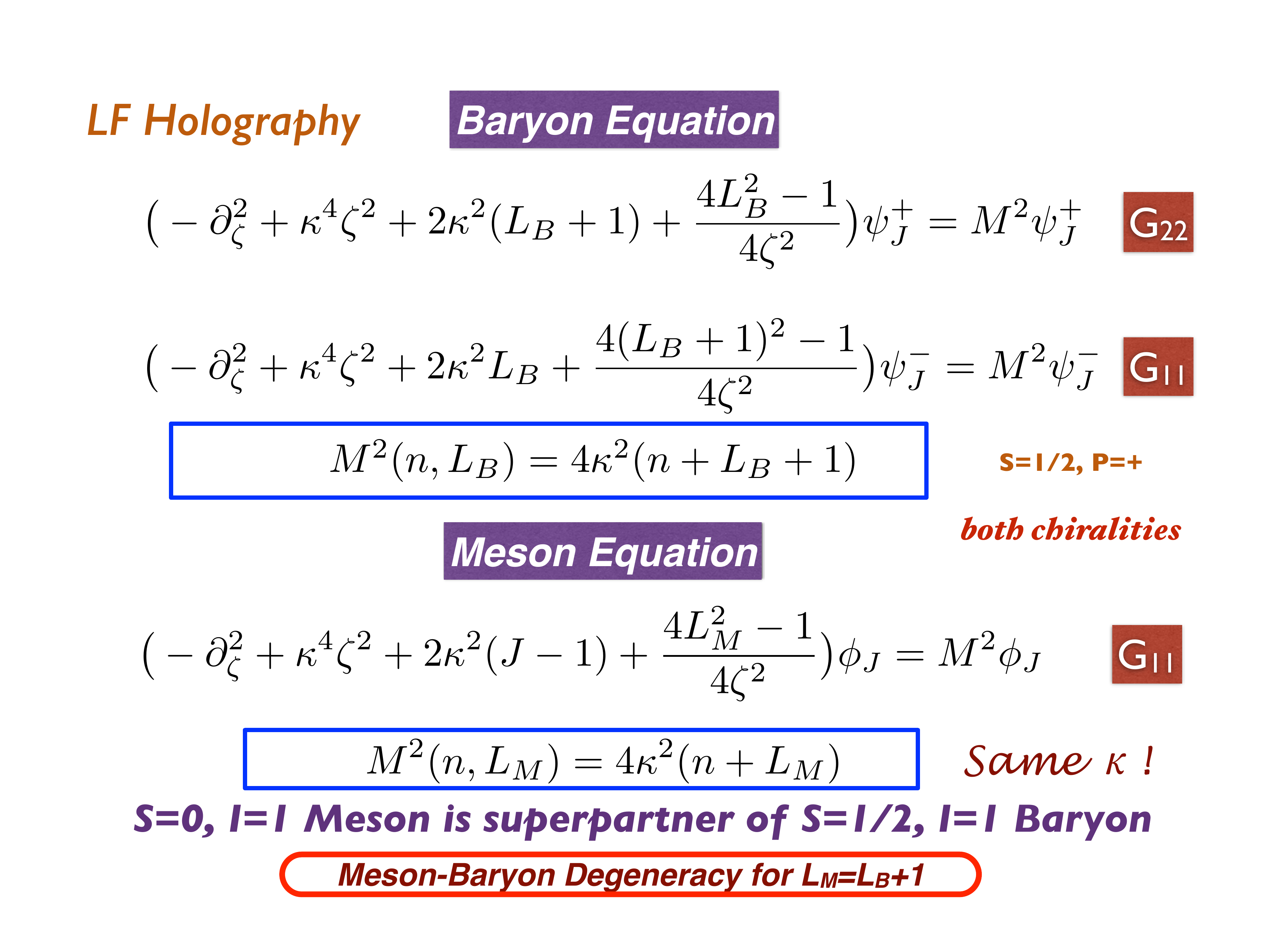}
\label{FigsJlabProc3}
\end{center}
\caption{The LF Schr\"odinger equations for baryons and mesons for zero quark mass derived from the Pauli $2\times 2$ matrix representation of superconformal algebra.  
The $\psi^\pm$  are the baryon quark-diquark LFWFs where the quark spin $S^z_q=\pm 1/2$ is parallel or antiparallel to the baryon spin $J^z=\pm 1/2$.   The meson and baryon equations are identical if one identifies a meson with internal orbital angular momentum $L_M$ with its superpartner baryon with $L_B = L_M-1.$
See Refs.~\cite{deTeramond:2014asa,Dosch:2015nwa,Dosch:2015bca}.}
\end{figure} 

Superconformal algebra leads to effective QCD light-front bound-state equations for both mesons and baryons~\cite{deTeramond:2014asa,Dosch:2015nwa,Dosch:2015bca}. 
The resulting set of bound-state equations for confined quarks are shown in Fig. 2. The supercharges connect the baryon and meson spectra  and their Regge trajectories to each other in a remarkable manner: the superconformal algebra  predicts that the bosonic meson and fermionic baryon masses are equal if one identifies each meson with internal orbital angular momentum $L_M$ with its superpartner baryon with $L_B = L_M-1$; the meson and baryon  superpartners  then have the same parity.  Since
$ 2+ L_M = 3 + L_B$, the twist-dimension of the meson and baryon superpartners are also the same.   Superconformal algebra thus explains the phenomenological observation that Regge trajectories  for both mesons and baryons have parallel slopes. 

\begin{figure}
\begin{center}
\includegraphics[height=10cm,width=15cm]{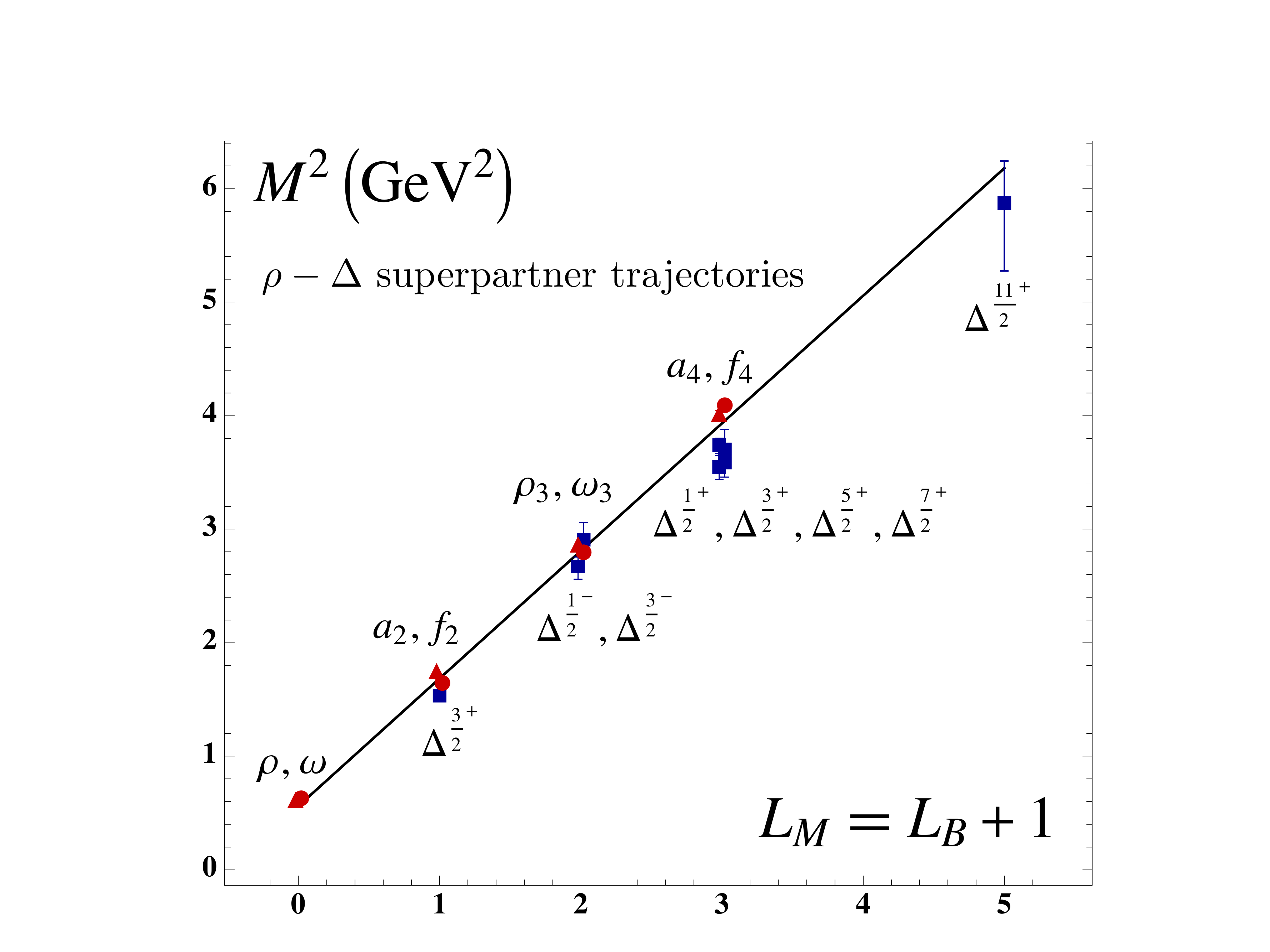}
\end{center}
\caption{Comparison of the $\rho/\omega$ meson Regge trajectory with the $J=3/2$ $\Delta$  baryon trajectory.   
Superconformal algebra  predicts the degeneracy of the  meson and baryon trajectories if one identifies a meson with internal orbital angular momentum $L_M$ 
with its superpartner baryon with $L_M = L_B+1.$
See Refs.~\cite{deTeramond:2014asa,Dosch:2015nwa}. }
\label{rhodelta.pdf}
\end{figure}

The comparison between the meson and baryon masses of the $\rho/\omega$ Regge trajectory with the spin-$3/2$ $\Delta$ trajectory 
is shown in Fig. \ref{rhodelta.pdf}.  The observed hadronic spectrum  with $N_C=3$  are seen to exhibit the supersymmetric features predicted by superconformal algebra.

\begin{figure}
\begin{center}
\includegraphics[height=10cm,width=15cm]{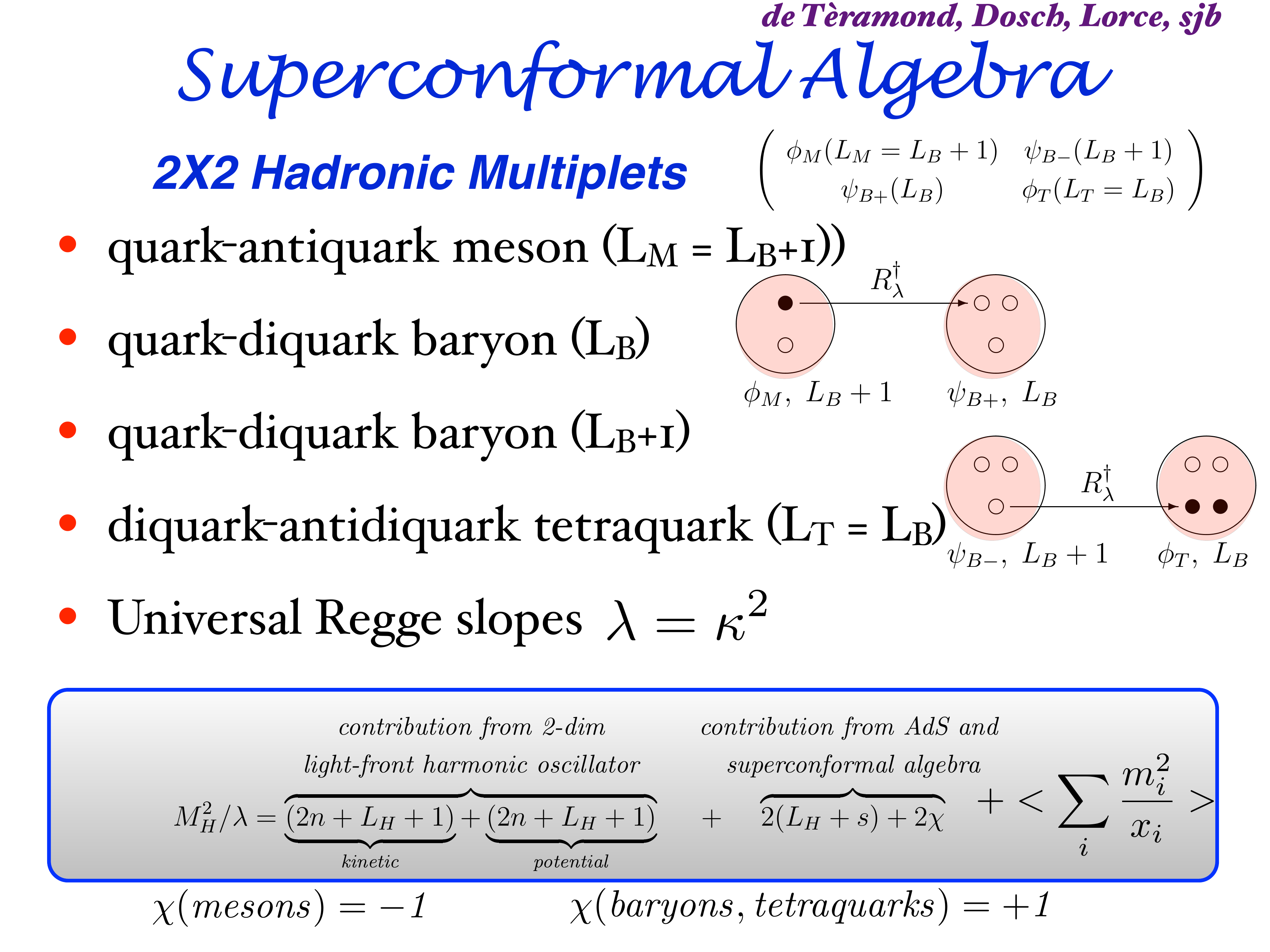}
\end{center}
\caption{The eigenstates of superconformal algebra have a $2 \times 2 $ representation of mass degenerate bosons and fermions:  a meson with $L_M=  L_B+1$, a baryon doublet with 
$L_B, L_B+1$ components and a tetraquark with $ L_T = L_B$. The breakdown of LF kinetic, potential, spin, and quark mass contributions to each hadron is also shown.  The virial theorem predicts the equality of the LF kinetic and potential contributions.
}
\label{2X2Multiplets}
\end{figure} 

As illustrated in fig. \ref{2X2Multiplets}, the hadronic eigensolutions of the superconformal algebra are   $2\times 2$ matrices connected internally by the supersymmetric algebra operators.
The eigensolutions of the supersymmetric conformal algebra thus have a $2 \times 2 $ Pauli matrix representation, where the upper-left component corresponds to mesonic $q \bar q$ color-singlet bound states, the two off-diagonal eigensolutions $\psi^{\pm}$  correspond to a pair of Fock components of baryonic quark-diquark states with equal weight, where the quark spin is parallel or antiparallel to the baryon spin, respectively.  The fourth component corresponds to diquark anti-diquark (tetraquark) bound states. The resulting frame-independent color-confining bound-state LF eigensolutions can be identified with the hadronic eigenstates of confined quarks  for $SU(3)$ color.    In effect, two of the quarks of the baryonic  color singlet $qqq$ bound state  bind to a color $\bar {3_C}$ diquark bound state, which then binds by the same color force to the remaining $3_C$ quark.  As shown by t'Hooft in a string model~\cite{tHooft:2004doe}, the $Y$ configuration of three quarks is unstable -- and reduces to the quark-diquark configuration.  The matching of the meson and baryon spectra is thus due to the fact that the same color-confining potential that binds two quarks to a diquark  also  binds a quark to an antiquark.

Note that the same slope controls the Regge trajectories of both mesons and baryons in both the orbital angular momentum $L$ and the principal quantum number $n$.
Only one mass parameter $\kappa = \omega^2$  appears; it sets the confinement scale and the hadron mass scale in the  chiral limit, as well as  the length scale which underlies hadron structure.  We will also use the notation $\lambda= \kappa^2$.    In addition to the meson and baryon eigenstates, one also predicts color-singlet {\it tetraquark}  diquark-antidiquark bound states with the same mass as the baryon.

The LF Schr\"odinger Equations for baryons and mesons derived from superconformal algebra  are shown  in Fig. 2.
As explained above, the baryons on the proton (Delta) trajectory are bound states of a quark with color $3_C$ and scalar (vector)  diquark with color $\bar 3_C$ 
The proton eigenstate labeled $\psi^+$ (parallel quark and baryon spins) and $\psi^-$ (anti parallel quark and baryon spins)  have equal Fock state probability -- a  feature of ``quark chirality invariance".  Predictions for the static properties of the nucleons are discussed in ref.~\cite{Liu:2015jna}.

Superconformal algebra also predicts that the LFWFs of the superpartners are related, and thus the corresponding elastic and transition form factors are identical.   The resulting  predictions for meson and baryon timelike form factors can be tested in $e^+ e^- \to H \bar H^\prime $ reactions. 

One can generalize these results to heavy-light $[\bar Q q] $ mesons and  $[Q [qq]]$ baryons~\cite{Dosch:2016zdv}.  The Regge slopes are found to increase for heavy $m_Q$ as expected from heavy quark effective field theory;  however, the supersymmetric connections between the heavy-light hadrons is predicted to be maintained.

The LFWFs thus play the same role in hadron physics as the Schr\"odinger wavefunctions which encode the structure of atoms in QED.  The elastic and transition form factors of hadrons, weak-decay amplitudes and distribution amplitudes are overlaps of LFWFs; structure functions, transverse momentum distributions
and other inclusive observables 
are constructed from the squares of the LFWFs.     In contrast one cannot compute form factors of hadrons or other current matrix elements of hadrons from overlap of the usual ``instant" form wavefunctions since one must also include contributions  where the photon interacts with connected but acausal vacuum-induced currents.
The calculation of deeply virtual Compton scattering using LFWFs is given in ref. \cite{Brodsky:2000xy}.  One can also compute the gravitational form factors of hadrons.  In particular, 
one can show that the anomalous gravitomagnetic moment $B(q^2=0)$ vanishes identically for any LF Fock state~\cite{Brodsky:2000ii}, in agreement with the equivalence theorem of gravity~\cite{Kobzarev:1962wt,Teryaev:1999su}.

\begin{figure}
\begin{center}
\includegraphics[height=10cm,width=15cm]{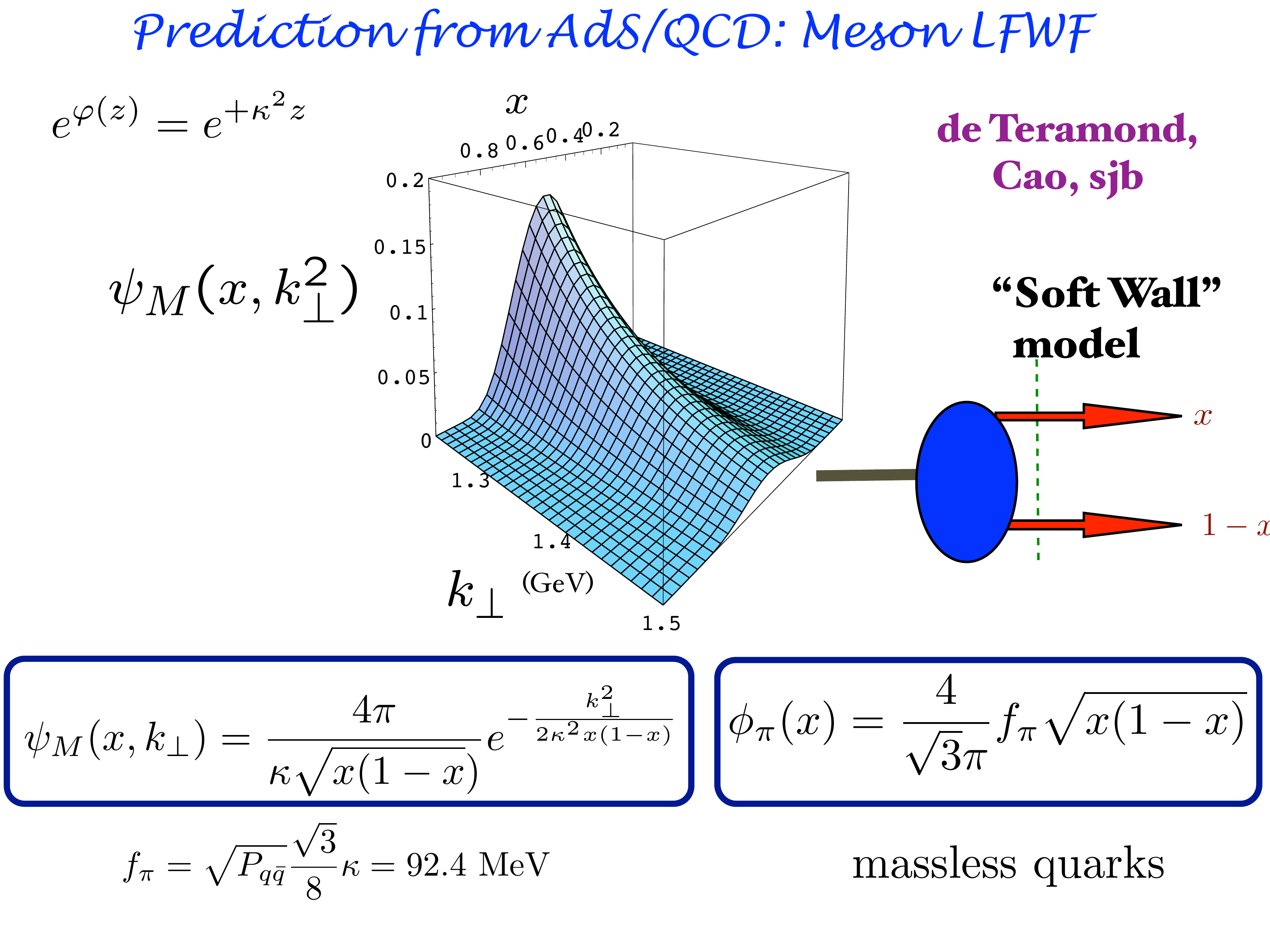}
\end{center}
\caption{Prediction from AdS/QCD and Light-Front Holography for  meson LFWFs  $\psi_M(x,\vec k_\perp)$   and the pion distribution amplitude.     
}
\label{MesonLFWF.pdf}
\end{figure}

The hadronic LFWFs predicted by light-front holography and superconformal algebra  are  functions of the LF kinetic energy $\vec k^2_\perp/ x(1-x)$ -- the conjugate of the LF radial variable $\zeta^2 = b^2_\perp x(1-x)$ -- times a function of $x(1-x)$; they do not factorize as a  function of $\vec k^2_\perp$ times a function of $x$.  The resulting  nonperturbative pion distribution amplitude $\phi_\pi(x) = \int d^2 \vec k_\perp \psi_\pi(x,\vec k_\perp) = (4/  \sqrt 3 \pi)  f_\pi \sqrt{x(1-x)}$,  see Fig. \ref{MesonLFWF.pdf}, which controls hard exclusive process, is  consistent with the Belle data for the photon-to-pion transition form factor~\cite{Brodsky:2011xx}.  
The AdS/QCD light-front holographic eigenfunction for the $\rho$ meson LFWF $\psi_\rho(x,\vec k_\perp)$ gives excellent 
predictions for the observed features of diffractive $\rho$ electroproduction $\gamma^* p \to \rho  p^\prime$,  as shown by Forshaw and Sandapen~\cite{Forshaw:2012im}

\section{Light-Front Holography} 
 
\begin{figure}
 \begin{center}
\includegraphics[height= 12cm,width=15cm]{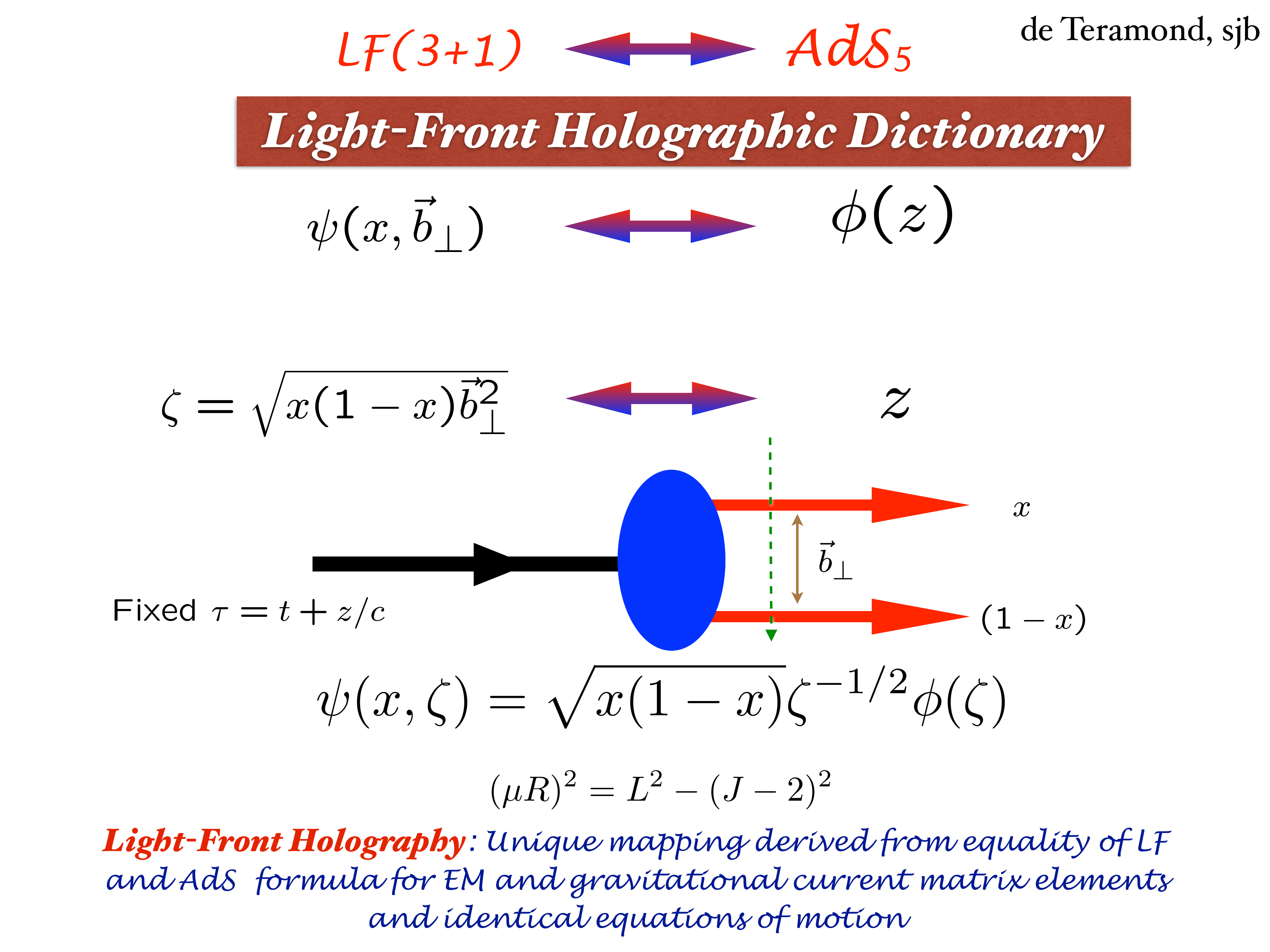}
\end{center}
\caption{The holographic dictionary which maps the fifth dimension variable $z$ of  the five-dimensional AdS$_5$ space to the LF radial variable $\zeta$ where 
$\zeta^2  =  b^2_\perp(1-x)$.   The same physics transformation maps the AdS$_5$  and $(3+1)$ LF expressions for electromagnetic and gravitational form factors to each other. 
From ref.~\cite{deTeramond:2013it}}
\label{dictionary}
\end{figure}

\begin{figure}
 \begin{center}
\includegraphics[height= 14cm,width=16cm]{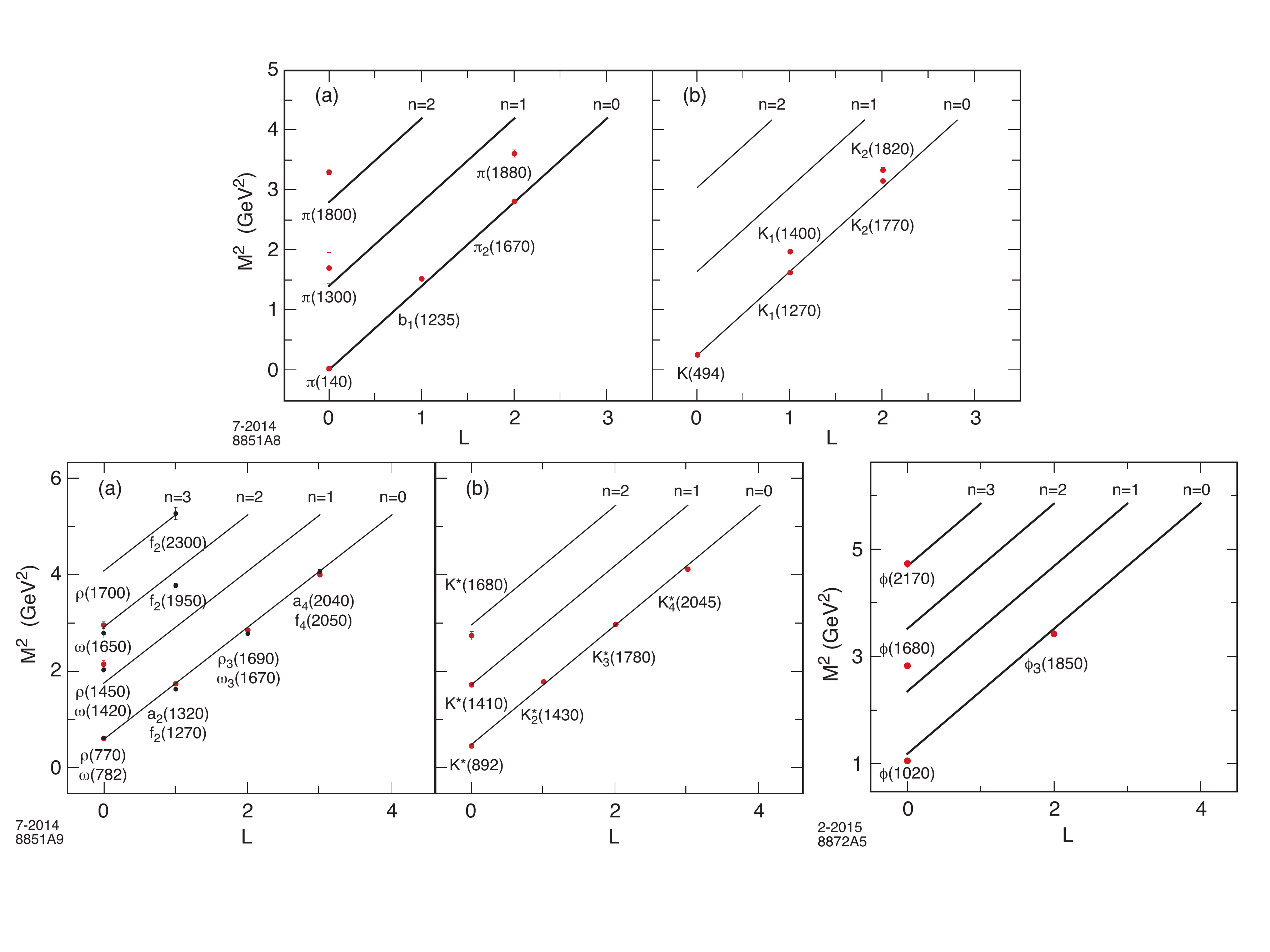}
\end{center}
\caption{Comparison of the AdS/QCD prediction  $M^2(n, L, S) = 4\kappa^2(n+L +S/2)$ for the orbital $L$ and radial $n$ excitations of the meson spectrum with experiment.   The pion is predicted to be massless for zero quark mass. The $u,d,s$ quark masses can be taken into account by perturbing in $<m_q^2/x>$.   The fitted value of $\kappa = 0.59$ MeV for pseudoscalar mesons, 
and  $\kappa = 0.54$ MeV  for vector mesons. }
\label{ReggePlot.pdf}
\end{figure}

Five-dimensional AdS$_5$ space provides a geometrical representation of the conformal group.
The color-confining light-front  equation for mesons of arbitrary spin $J$ can be derived~\cite{deTeramond:2013it}
from the holographic mapping of  the ``soft-wall model" modification of AdS$_5$ space for the specific dilaton profile $e^{+\kappa^2 z^2},$  where one identifies the fifth dimension coordinate $z$ with the light-front coordinate $\zeta$.  
Remarkably ,  AdS$_5$  is holographically dual to $3+1$  spacetime at fixed light-front time $\tau = t+ z/c$.  
The holographic dictionary is summarized in Fig. \ref{dictionary} 
The combination of light-front dynamics, its holographic mapping to AdS$_5$ space, and the dAFF procedure provides new insight into the physics underlying color confinement, the nonperturbative QCD coupling, and the QCD mass scale.  A comprehensive review is given in  Ref.~\cite{Brodsky:2014yha}.  The $q \bar q$ mesons and their valence LF wavefunctions are the eigensolutions of the frame-independent relativistic bound state LF Schr\"odinger equation -- the same meson equation that is derived using superconformal algebra.
The mesonic $q\bar  q$ bound-state eigenvalues for massless quarks are $M^2(n, L, S) = 4\kappa^2(n+L +S/2)$.
The equation predicts that the pion eigenstate  $n=L=S=0$ is massless at zero quark mass. The  Regge spectra of the pseudoscalar $S=0$  and vector $S=1$  mesons  are 
predicted correctly, with equal slope in the principal quantum number $n$ and the internal orbital angular momentum $L$.  A comparison with experiment is shown in Fig. \ref{ReggePlot.pdf}.

Light-Front Holography  not only predicts meson and baryon  spectroscopy  successfully, but also hadron dynamics, including  vector meson electroproduction,  hadronic light-front wavefunctions, distribution amplitudes, form factors, and valence structure functions.   The  application to the deuteron elastic form factors and structure functions is given 
in ref.~\cite{Gutsche:2015xva,Gutsche:2016lrz}

\section{Color Confinement  from LF Holography}

Remarkably, the light-front potential using the dAFF procedure has the unique form of a harmonic oscillator $\kappa^4 \zeta^2$ in the 
light-front invariant impact variable $\zeta$ where $ \zeta^2Ê = b^2_\perp x(1-x)$. The result is  a single-variable frame-independent relativistic equation of motion for  $q \bar q $ bound states, a ``Light-Front Schr\"odinger Equation"~\cite{deTeramond:2008ht}, analogous to the nonrelativistic radial Schr\"odinger equation in quantum mechanics. The same result, including spin terms, is obtained using  light-front holography  -- the duality between the front form and AdS$_5$, the space of isometries of the conformal group -- if one  
modifies the action of AdS$_5$ by the dilaton $e^{\kappa^2 z^2}$ in the fifth dimension $z$.  The  Light-Front Schr\"odinger Equation  incorporates color confinement and other essential spectroscopic and dynamical features of hadron physics, including a massless pion for zero quark mass and linear Regge trajectories with the same slope  in the radial quantum number $n$   and internal  orbital angular momentum $L$.      
When one generalizes this procedure using superconformal algebra, the resulting light-front eigensolutions predict a unified Regge spectroscopy of meson, baryon, and tetraquarks, including remarkable supersymmetric relations between the masses of mesons and baryons of the same parity.

It is interesting to note that the contribution of the {\it `H'} diagram to $Q \bar Q $ scattering is IR divergent as the transverse separation between the $Q$  
and the $\bar Q$ increases~\cite{Smirnov:2009fh}.  This is a signal that pQCD is inconsistent without color confinement.  The sum of such diagrams could sum to the confinement potential $\kappa^4 \zeta^2 $ dictated by the dAFF principle that the action remains conformally invariant despite the appearance of the mass scale $\kappa$ in the Hamiltonian.
The $\kappa^4 \zeta^2$ confinement interaction between a $q$ and $\bar q$ will induce a $\kappa^4/s^2$ correction to $R_{e^+ e^-}$, replacing the $1/ s^2$ signal usually attributed to a vacuum gluon condensate.

\begin{figure}
 \begin{center}
\includegraphics[height= 12cm,width=15cm]{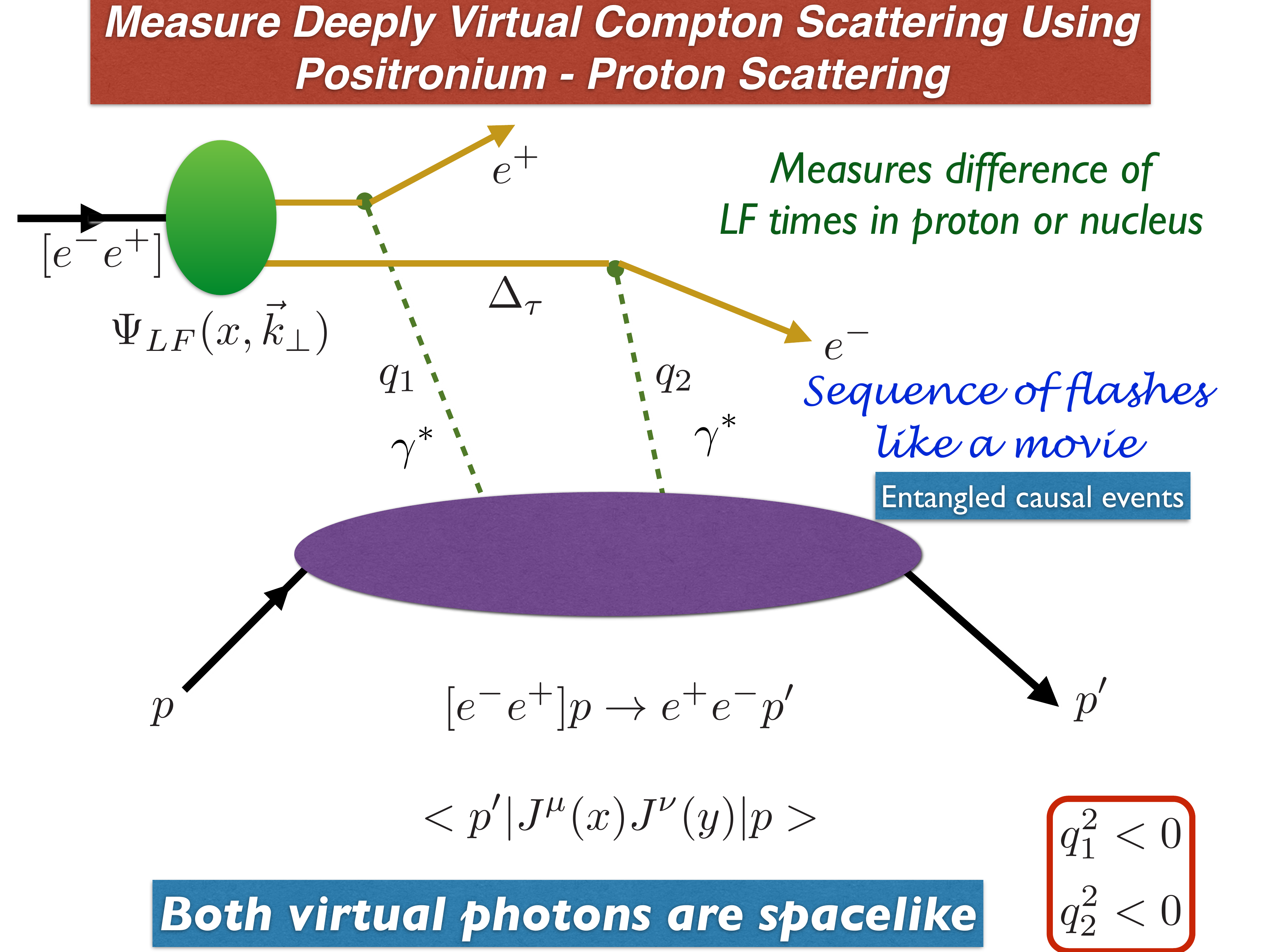}
\end{center}
\caption{Doubly Virtual Compton scattering on a proton (or nucleus) can be measured for two {\it spacelike} photons $q^2_1, q^2_2 <0$ 
with minimal, tunable, skewness $\xi$ using positronium-proton scattering $[e^+ e^-] p \to e^+ e^- p'$.  
One can also measure double deep inelastic scattering and elastic positronium-proton scattering.  
One can also create a beam of  ``true muonium" atoms $[\mu^- \mu^-]$~\cite{Brodsky:2009gx,Banburski:2012tk} using Bethe-Heitler pair production just below threshold.
 }
\label{Positronium}
\end{figure}

It should be emphasized that the value of the mass scale $\kappa$ is not determined in an absolute sense by QCD.    Only ratios of masses are determined, and the theory has dilation invariance under $\kappa \to C \kappa $.    In a sense, chiral QCD has an ``extended conformal invariance."  The resulting new time variable  which retains the conformal invariance of the action, has finite support, conforming to the fact that the LF time between the interactions with the confined constituents is finite.  
The finite time difference $\Delta \tau$ between the LF times  $\tau$ of the quark constituents of the proton could be measured using positronium-proton scattering $[e^+ e^-] p \to e^+ e^- p'$.  This process, which measures double diffractive deeply virtual Compton scattering for two spacelike photons, is illustrated in Fig.~\ref{Positronium}.  One can also study the dissociation of relativistic positronium
atoms to an electron and positron with light front momentum fractions $x$ and $1-x$ and  opposite transverse momenta in analogy to the E791 measurements of the diffractive dissociation of the pion to two jets~\cite{Ashery:2000yj}.
The LFWF of positronium in the relativistic domain is the central input.
One can produce a relativistic positronium beam  using the collisions of laser photons with high energy photons or by 
using Bethe-Heitler pair production below the $e^+ e^-$ threshold.
The production of parapositronium via the collision of photons is analogous to pion production in two-photon interactions and Higgs production via gluon-gluon fusion.

\section{Light-Front Theory and QCD}

One can derive the exact form of the light-front Hamlitonian $H_{LF}$ directly from the QCD Lagrangian and avoid ghosts and longitudinal  gluonic degrees of freedom by choosing the light-cone gauge  $A^+ =0$.  
Quark masses appear in the LF kinetic energy as $\sum_i {m^2\over x_i}$. This can be derived from the Higgs theory quantized using LF dynamics~\cite{Srivastava:2002mw}.
The confined quark field $\psi_q$ couples to the background Higgs field  $g_{\overline \Psi_q } <H >  \Psi_q$ via its Yukawa  scalar matrix element  coupling  
$g_q <H > \bar u(p) 1 u(p) = m_q \times {m_q \over x} = {m^2\over x}.$  The usual  Higgs vacuum expectation value  $<H > $ is replaced by a constant zero mode when one quantizes the Standard Model using light-front quantization~\cite{Srivastava:2002mw}.

PQCD factorization theorems and  the DGLAP  \cite{Gribov:1972ri,Altarelli:1977zs,Dokshitzer:1977sg} and ERBL \cite{Lepage:1979zb,Lepage:1980fj,Efremov:1979qk,Efremov:1978rn} evolution equations can also be derived using the light-front Hamiltonian formalism~\cite{Lepage:1980fj}.  In the case of an electron-ion collider, one can represent the cross section for $e-p$ colisions as a convolution of the hadron and virtual photon structure functions times the subprocess cross-section in analogy to hadron-hadron colisions.   This nonstandard description of $\gamma^* p \to X$ reactions  gives new insights into electroproduction physics -- physics not apparent   in the usual infinite-momentum frame description, such as the dynamics of heavy quark-pair production.  
Intrinsic heavy quarks at high $x$  also play an important role~\cite{Brodsky:2015uwa}.

The LF Heisenberg equation can in principle be solved numerically by matrix diagonalization  using the ``Discretized Light-Cone  Quantization" (DLCQ)~\cite{Pauli:1985pv} method.  Anti-periodic boundary conditions in 
$x^-$ render the $k^+$ momenta  discrete  as well as  limiting the size of the Fock basis.   In fact, one can easily solve $1+1 $ quantum field theories such as QCD$(1+1)$~\cite{Hornbostel:1988fb} for any number of colors, flavors and quark masses using DLCQ. 
Unlike lattice gauge theory, the nonpertubative DLCQ analysis is in Minkowski space, is frame-independent, and is free of fermion-doubling problems.   
AdS/QCD, based on the  AdS$_5$ representation of the conformal group in five dimensions, maps to physical 3+1 space-time at fixed LF time;  this correspondence, ``light-front holography"~\cite{deTeramond:2008ht},  is  now providing a  color-confining approximation to $H_{LF}^{QCD}$ for QCD(3+1).  This method gives a remarkable first approximation to  hadron spectroscopy and  hadronic LFWFs.   A new method for solving nonperturbative QCD ``Basis Light-Front Quantization" (BLFQ)~\cite{Vary:2014tqa},  uses the eigensolutions of a color-confining approximation to QCD (such as LF holography) as the basis functions,  rather than the plane-wave basis used in DLCQ, thus incorporating the full dynamics of QCD.  LFWFs can also be determined from the covariant Bethe-Salpeter wavefunction by integrating over $k^-$~\cite{Brodsky:2015aia}.  
A review of the light-front formalism is given in Ref.~\cite{Brodsky:1997de}.

\section{Measuring LFWFs of Hadrons, Atoms, and Nuclei}

One can in fact measure the LFWFs of QED atoms using diffractive dissociation.

For example, suppose one creates a relativistic positronium beam.   It will dissociate by Coulomb exchange in a thin target:
$[e^+ e^- ] + Z \to e^+ e^- Z$.
The momentum distribution of the leptons in  the LF variables $x$ and $k_\perp$ will determine   the first derivative of the atomic LFWF 
${d\over d k_\perp} \psi(x,\vec k_\perp)$.
When ${k^2_\perp \over x(1-x)} > 4 m^2_e$ one can observe the transition from NR Schr\"odinger theory  where $ \psi(x,\vec k_\perp) \propto {1\over k^4_\perp}$
to the relativistic domain, where $ \psi(x,\vec k_\perp) \propto {1\over k^2_\perp}$.
One can thus test predictions from BLFQ (Basis Light-Front Quantization)~\cite{Vary:2013kma}.  Higher Fock states are also possible, such as  $[e^+ e^-] + Z \to e^+ e^-  \gamma Z$ and 
$[e^+ e^-] + Z \to e^+ e^-  e^+ e^- Z$.?

Positronium dissociation is analogous to the Ashery measurements of the pion LFWF:
$\pi A \to Jet Jet A$ ~\cite{Ashery:2005wa}, where one observes the transition from Gaussian fall-off to power law  fall-off at large $1\over k^2_\perp $ as predicted by AdS/QCD.
When ${k^2_\perp \over x(1-x)} > 4 m^2_e$ one can measure the transition from NR Schr\"odinger theory to the relativistic domain, where $ \psi(x,\vec k_\perp) \propto {1\over k^2_\perp}$. 
Similarly, one could also measure the LFWF of a nucleus like a deuteron by dissociating relativistic ions $d A \to p n A$ .   At large $1\over k^2_\perp $ one  could observe the transition to the ``hidden-color" Fock states predicted by QCD~\cite{Brodsky:1983vf}.

\section{Calculations using LF-Time-Ordered Perturbation Theory and Hadronization at the Amplitude Level}

LF-time-ordered perturbation theory  can be advantageous  for perturbative QCD calculations.  
An excellent example  of LF-time-ordered perturbation theory is the computation of multi-gluon scattering amplitudes by Cruz-Santiago and Stasto~\cite{Cruz-Santiago:2015dla}.  
In this method, the propagating particles  are on their respective mass shells:  $k_\mu k^\mu = m^2$, and intermediate states are off-shell in invariant mass;  {\it i.e.}, $P^- \ne \sum k^-_i$.  Unlike instant form, where one must sum  $n!$ frame-dependent  amplitudes, only  the $\tau$-ordered diagrams where each propagating particle has  positive $k^+ =k^0+k^z$  can contribute. 
The number of nonzero amplitudes is also greatly reduced by noting that the total angular momentum projection $J^z = \sum_i^{n-1 } L^z_i + \sum^n_i S^z_i$ and the total $P^+$ are  conserved at each vertex.  In a renormalizable theory, the change in orbital angular momentum is limited to $\Delta L^z =0,\pm 1$ at each vertex~\cite{Chiu:2017ycx}

A remarkable advantage of LF time-ordered perturbation theory (LFPth) is that the calculation of a subgraph of any order in pQCD only needs to be done once;  the result can be stored in a ``history" file.  This is due to the fact that in LFPth the numerator algebra is independent of the process; the denominator changes, but only by a simple shift of the initial $P^-$.   Another simplification is that loop integrations are three dimensional: $\int d^2\vec k_\perp \int^1_0 dx.$   Unitarity  and explicit
renormalization can be implemented using the ``alternate denominator" method which defines the required subtraction counterterms~\cite{Brodsky:1973kb}.
 
The new insights into color confinement given by AdS/QCD suggest that one could compute ``hadronization at  the amplitude level"~\cite{Brodsky:2009dr} using  the confinement interaction and the LFWFs predicted by  AdS/QCD and Light-Front Holography. 
For example, as illustrated in fig. \ref{had1},  the meson LFWF connects  the off-the-invariant mass shell quark and antiquark to the on-shell asymptotic physical meson state.

The invariant mass of a color-singlet cluster ${\cal M}$  is the key variable which separates perturbative and nonperturbative dynamics.
For example, consider $e^+ e^- $ annihilation using LF $\tau$ - ordered perturbation theory.   
At an early stage in LF time,  the annihilation will produce jets of quarks and gluons in an intermediate state off the $P^-$ energy shell. 
If a color-singlet cluster of partons in a jet satisfies ${\cal M}^2 -M^2_H <  \kappa^2  $,
the cluster constituents are effective degrees of freedom will be ruled by the $\kappa^4\zeta^2$ color-confinement potential.
At this stage, the LFWF $\psi_H$ converts the off-shell partons to 
the asymptotic states, the on-shell hadrons.  If ${\cal M}^2 > \kappa^2$ one can apply pQCD corrections; e.g. from DGLAP and ERBL evolution~\cite{Lepage:1979zb,Lepage:1980fj,Efremov:1979qk,Efremov:1978rn} .

\begin{figure}
 \begin{center}
 \includegraphics[height= 9cm,width=15cm]{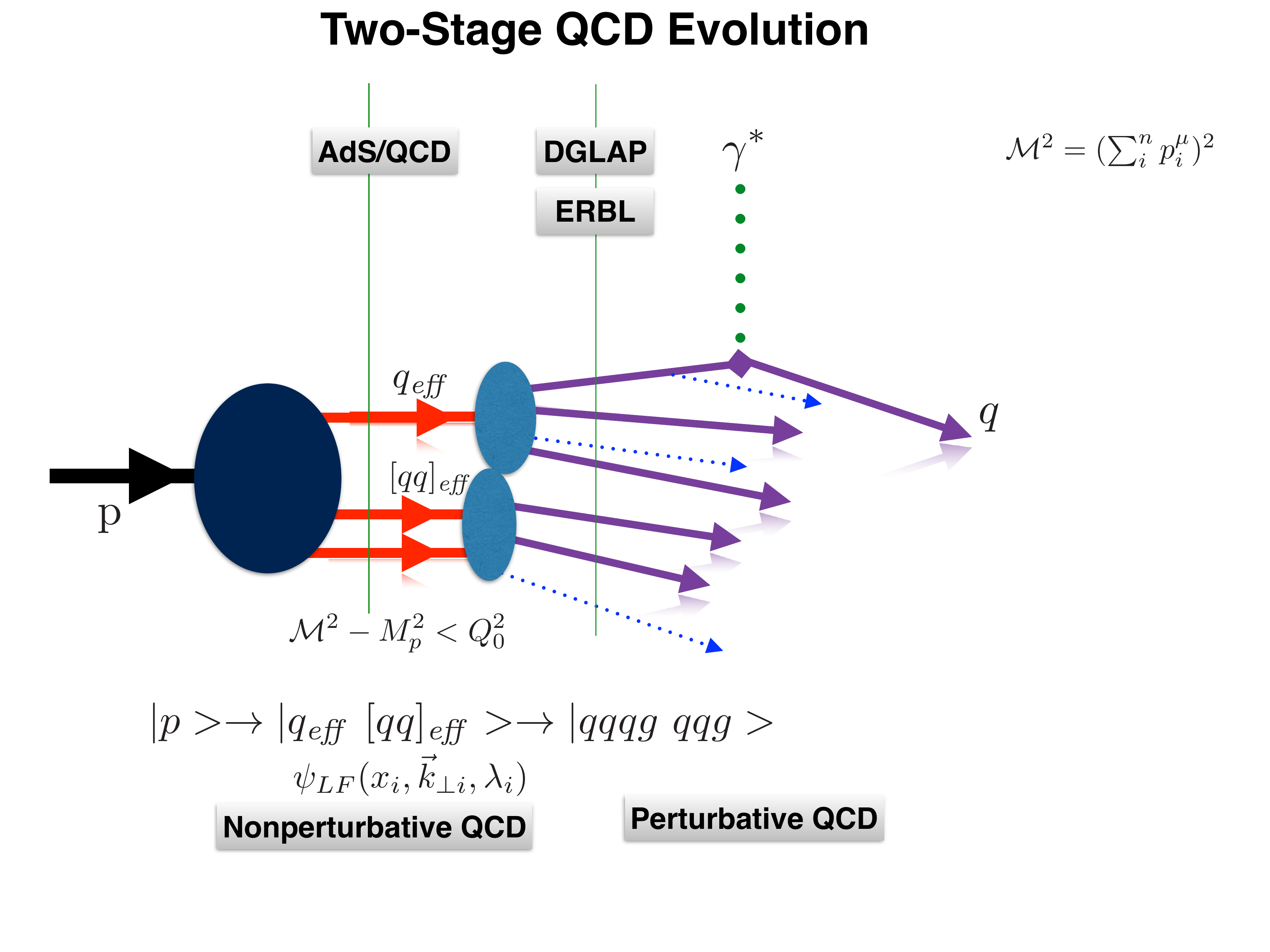}
\includegraphics[height=  9cm,width=15cm]{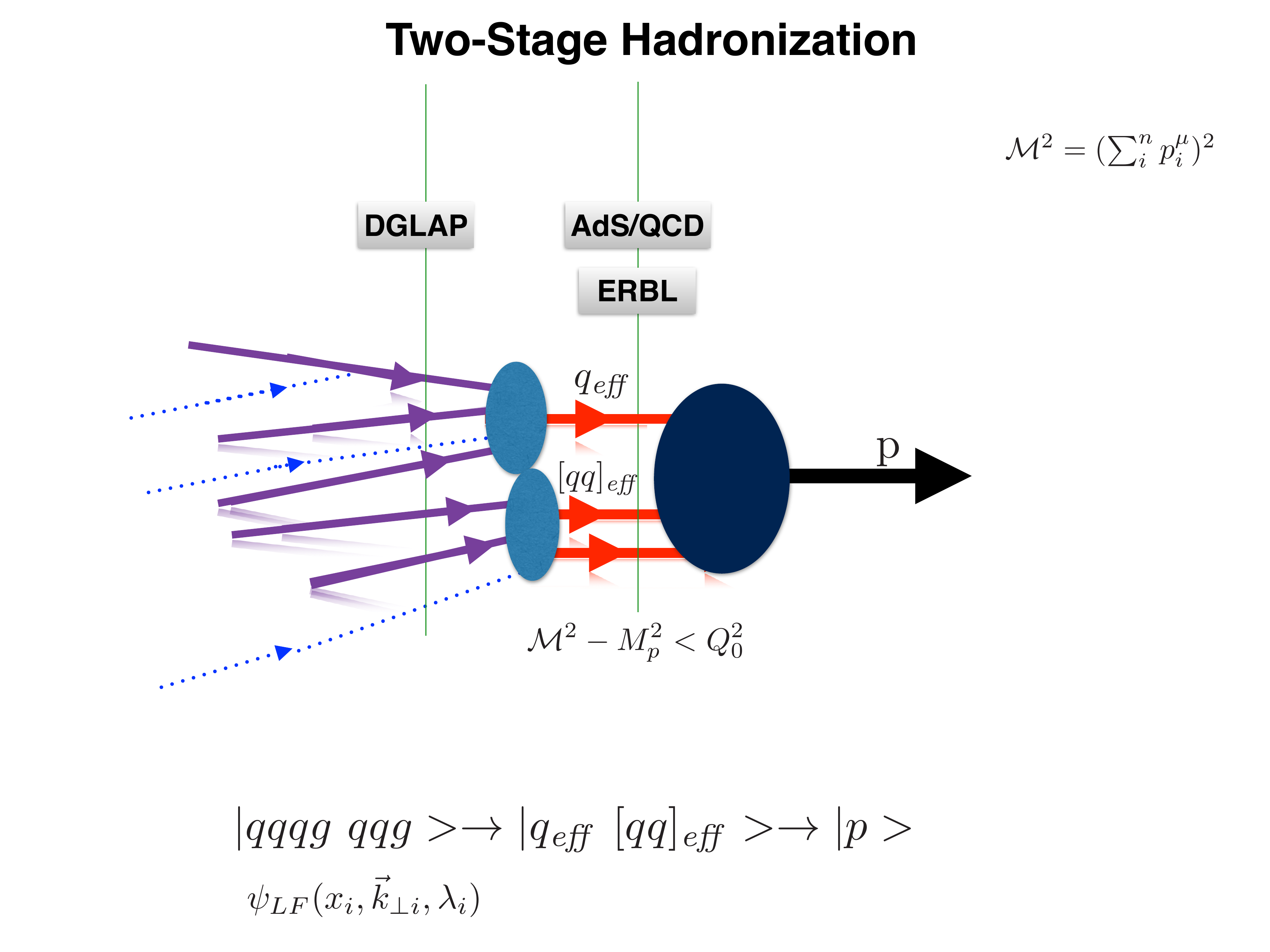}
\end{center}
\caption{  {\it A}. A model for evolution starting with  a nonperturbative hadronic LFWF.  {\it B}. Hadronization and evolution ending with a hadronic  LFWF.  The intermediate quark and gluon states are off the $P^-$ energy shell and thus off-the-invariant mass shell  ${\cal M}^2  > m^2_H$  
In the off-shell domain ${\cal M}^2 -M^2_H >  \kappa^2,  $ the intermediate quarks and gluons obey the DGLAP and ERBL QCD evolution.   
If  a cluster of quarks and antiquarks satisfies ${\cal M}^2 -M^2_H <  \kappa^2$, the intermediate state sees the color confinement interaction.
The meson LFWF connects the intermediate $q \bar q $ state, which is  off of the $P^-$ energy shell and thus off-the-invariant mass shell  ${\cal M}^2  > m^2_H$ to the  physical meson state with 
${\cal M}^2  = m^2_H$.  The LF angular momentum $J^z$ is conserved at every vertex. }
\label{Had3}
\end{figure}

A model for the two stages of hadronization and evolution is illustrated in fig. \ref{Had3}.  In the off-shell domain ${\cal M}^2 -M^2_H >  \kappa^2,  $ the intermediate quarks and gluons obey DGLAP and ERBL evolution. 

Thus quarks and gluons can appear in intermediate off-shell states, but only hadrons are produced asymptotically.   
Thus the AdS/QCD Light-Front Holographic model  suggests how one can implement the transition between perturbative and nonperturbative QCD.  For a QED analog, see  Refs.~\cite{Brodsky:2009gx,Banburski:2012tk}. 

\section{Light-Front Spin and Light-Front $J^z$ Conservation}
 
A central, unique property of light-front quantization is $J^z$ conservation~\cite{Chiu:2017ycx}; the $z$-component of  angular momentum remains unchanged under Lorentz transformations generated by the light-front kinematical boost operators.   The spin along
the $\hat z$ direction defined by the light-front Lorentz transformation is preserved because $<J^3>_{LF}= s^z$ for all momenta $p^\mu$.  $J^z$ conservation underlies the Jaffe spin sum rule~\cite{Jaffe:1987sx}.

Particles in the front form move with positive $k^+ = k^0+k^z \ge 0.$ 
The quantization axis for $J^z$ for each particle is the same axis $\hat z$ which defines LF time $\tau = t + z/c$.
Thus $S^z$ and $L^z$ refer to angular momentum in the $\hat z$ direction. 
As in nonrelativistic quantum mechanics,  $J^z =  \sum^n_{i=1} S^z_i + \sum^{n-1}_{n=1}  L^z_i$ for any $n$- particle intermediate or Fock state.  There are $n-1$ relative orbital angular momenta.   It is conserved at every vertex and is conserved overall for any process and ``LF helicity" refers to the spin projection $S^z$ of each particle and ``LF chirality"  is the spin projection $S^z$  for massless particles.   In a renormalizable theory $L^z$ can only change  by one unit at any vertex.  This leads to a rigorous selection rule for amplitudes at fixed order in pQCD~\cite{Chiu:2017ycx}: $|\Delta L^z |  \le n$ in an
$n$-th order perturbative expansion.  This  selection rule for the orbital angular momentum  can be used to eliminate interaction vertices in QED and QCD and provides an upper bound on the change of orbital angular momentum in scattering processes at any fixed order in perturbation theory.

By definition, spin and helicity can be used interchangeably in the front form.  
LF chirality is conserved by the vector current in electrodynamics and the axial current of electroweak interactions.   Each coupling  conserves quark chirality when the quark mass  is  set to zero.    A compilation of LF spinor matrix elements is given in ref.~\cite{Lepage:1980fj}.

Light-front spin is not the same  as the usual ``Wick helicity",  where spin is defined as the projection the particle's three-momentum $\vec k$.  
Wick helicity is thus not conserved unless all particles move in the same direction. Wick helicity can be frame dependent. 
For example, In the case of  $gg \to H$, the  Wick helicity assignment  is $(+1) + (+1) \to 0$ in the CM frame,  but it is 
$(+1) + (-1) \to 0$  for collinear gluons if the  two gluons move in the same direction.

The twist of a hadronic interpolating operator corresponds to the number of fields plus $|L^z|$.   The  pion LF Fock state  for  $\pi \to q \bar q$ with  twist-2 corresponds to $(J^z_\pi=0) \to (S^z_q =+ {1\over 2} ) +  (S^z= - {1\over 2} )$ with zero relative orbital angular momentum $L^z_{q \bar q}$.   
This is the Fock state of the pion that decays to $\ell \nu$ via the LF chiral-conserving axial current $\gamma^\mu \gamma_5$. 
The twist-3 pion in the OPE corresponds to 
$J^z_\pi =0 \to (S^z_q =+ {1\over 2} ) +  (S^z_{\bar q}= + {1\over 2} ) + (L^z =-1) $  or $ J^z=0 \to (S^z_q = - {1\over 2} ) +  (S^z_{\bar q}= - {1\over 2} ) + (L^z_{q \bar q} =-1), $
where $L^z$ is the relative orbital angular momentum between the quark and antiquark.  The twist-3 Fock state couples the pion to the chiral-flip pseudoscalar $\gamma_5$  operator.
The GMOR relation connects the twist-2 and twist-3 Fock states when $m_q \ne 0$~\cite{Brodsky:2012ku}. 
The twist-3 proton  with $J^z_p=+{1\over 2}$ in AdS/QCD is a bound state of a quark  with $S^z_p={1\over 2}$ and a spin-zero diquark $[qq]$  with $L^z_{q [qq]} = 0$, and the twist-4 proton in AdS/QCD is a bound state of a quark with $S^z_p=-{1\over 2}$ and spin-zero diquark $[qq]$  with relative orbital angular momentum $L^z_{q [qq]} = +1)$.  LF holography predicts equal probability for the twist-3 and twist-4 Fock states in the nucleon for $m_q=0.$

\begin{figure}
 \begin{center}
 \includegraphics[height= 8cm,width=15cm]{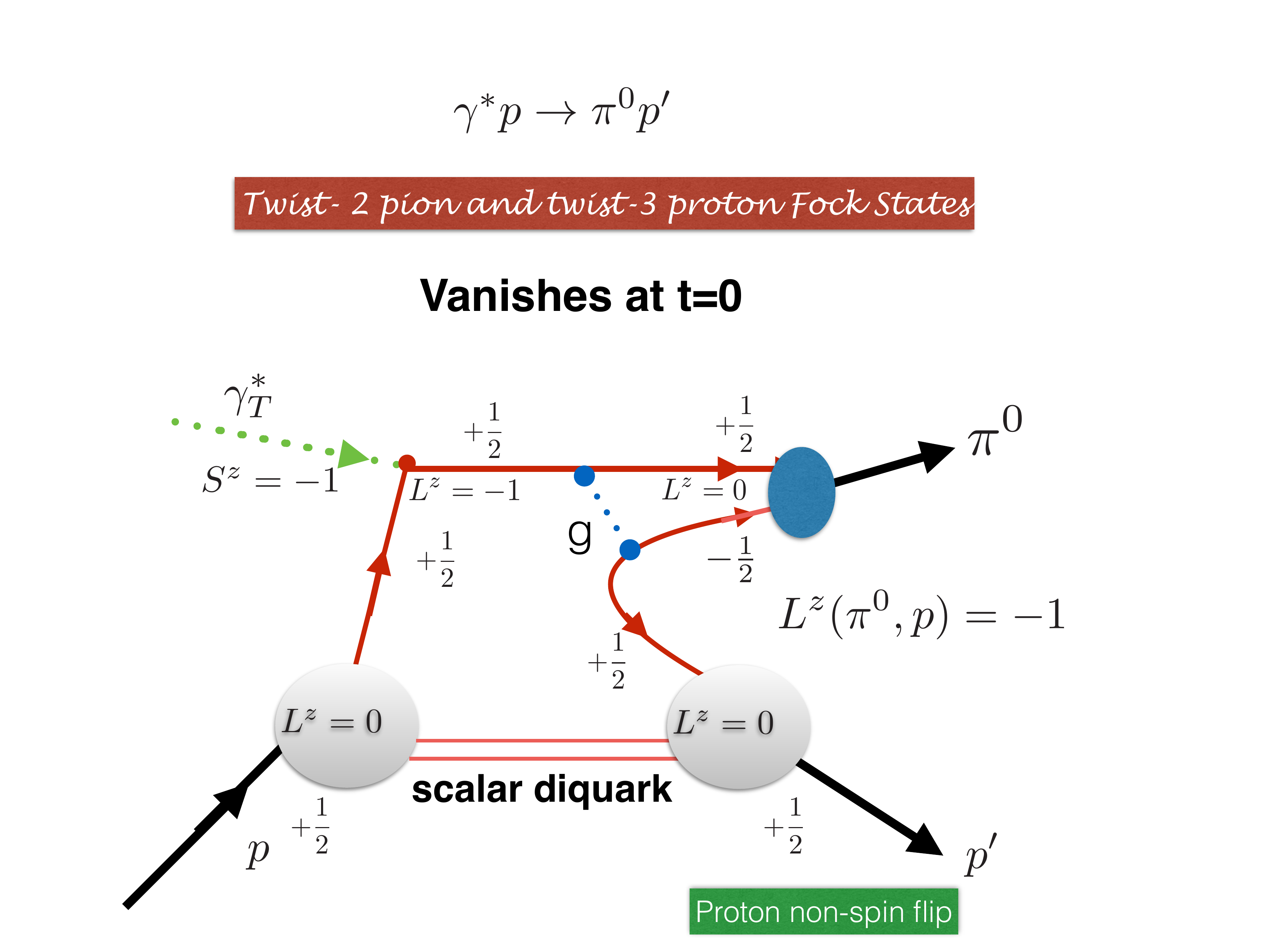}
\includegraphics[height=  8cm,width=15cm]{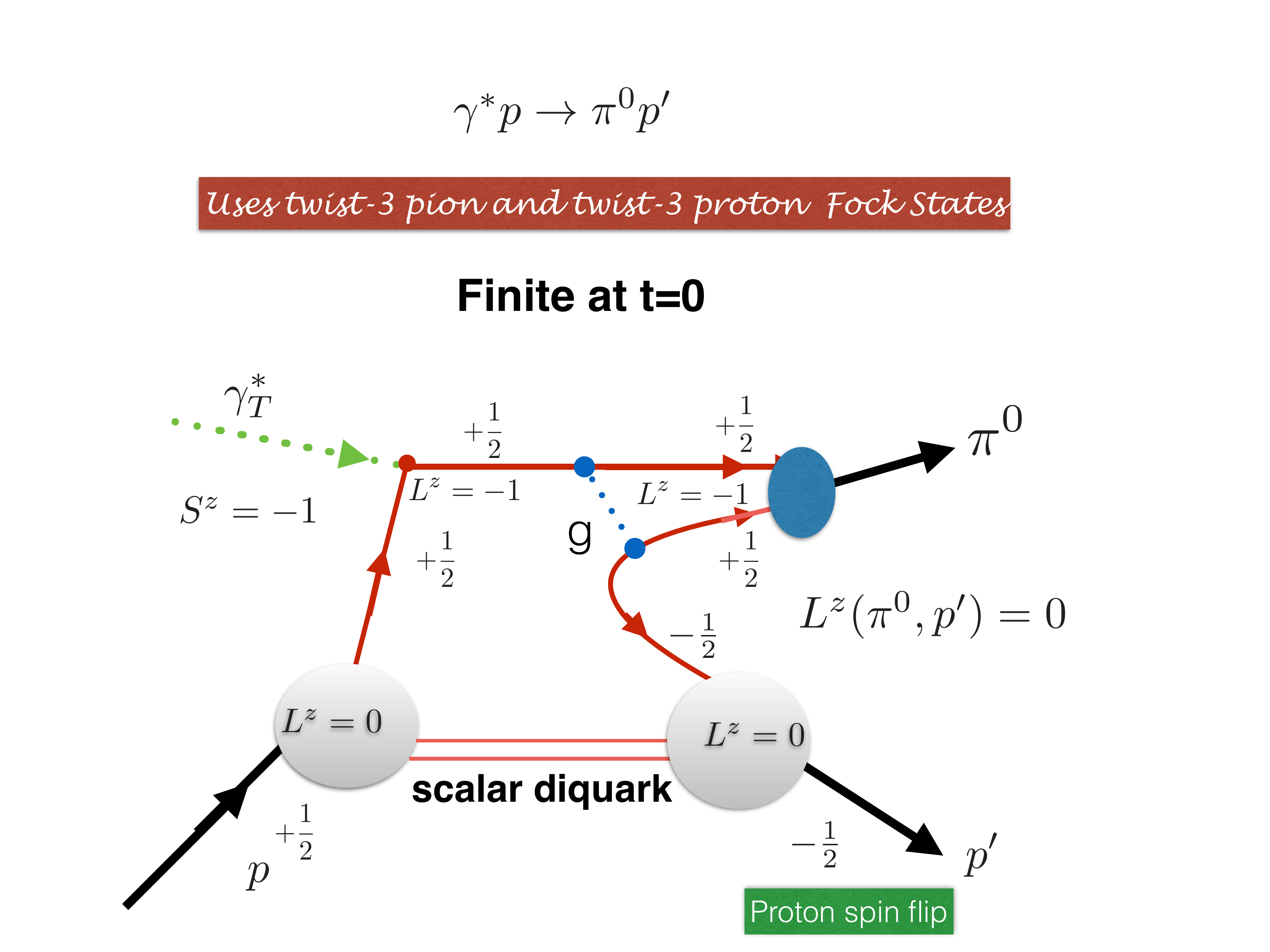}
\includegraphics[height=  8cm,width=15cm]{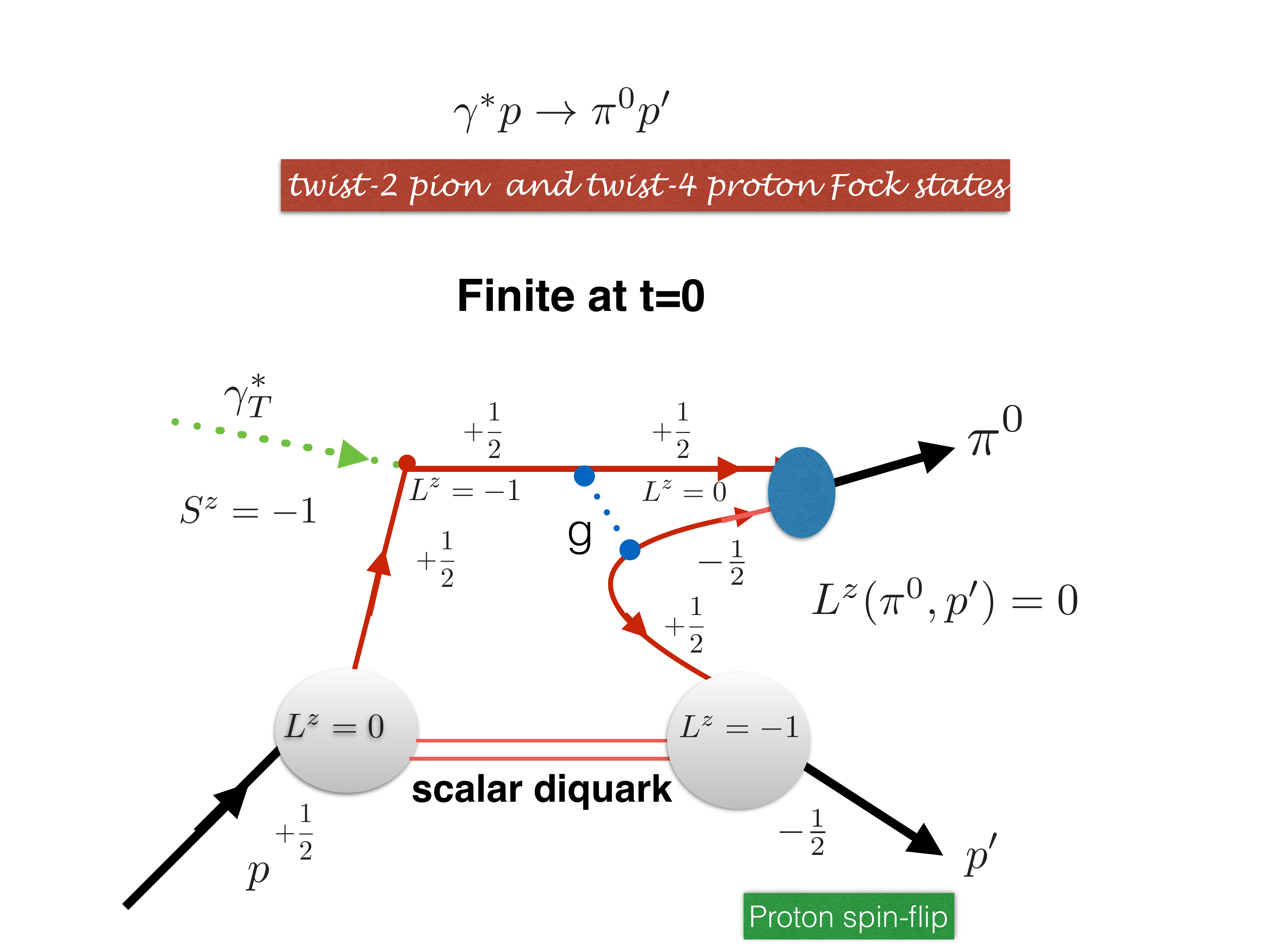}
\end{center}
\caption{Illustration of spin flow in $\gamma^* p \to \pi^0 p$}
\label{piontwist}
\end{figure}

One can  use LF $J^z$ conservation to determine the contribution of Fock states of different twist in a scattering amplitude by using the fact that amplitudes with nonzero relative $L^z$ between the outgoing particles vanish at in the forward direction.
For example, consider pion electroproduction 
$\gamma^* p \to \pi^0 p$ for a polarized photon with LF spin $S^z_\gamma= -1.$   If the proton's LF spin $S^z_p=-{1\over 2}$ is unchanged, 
then $J^z_{tot}= +{1\over 2}:  \gamma^*_T(S^z_\gamma = +1) + (S^z_p = - {1\over 2} )  \to   + (J^z_\pi =0)  + (S^z_p = - {1\over 2} ) +( L^z_{\pi^0 - p}=+1)  $  
vanishes at $t =0$ for the twist-2 pion.   However,  the non-spin-flip proton amplitude: 
$J^z_{tot}= {1\over 2}:  \gamma^*_T(S^z_\gamma = +1) + S^z_p = (- {1\over 2} )  \to  
[(S^z_q  = - {1\over 2})  +  (S^z_{\bar q}  = - {1\over 2} )  + ( L^z_{q \bar q}= +1) ] _{\pi^0} +  S^z_p = (- {1\over 2} ) $  for the  twist-3 pion Fock state is finite at $t = 0$. 
A similar result holds for the contribution of the twist-2 pion and twist-4 proton.
See  fig. \ref{piontwist}.

Similarly one can utilize the behavior of the the amplitude $\gamma^* He^4 \to \pi^0 He^4$  on a spin-zero helium target.
The pion twist-2 amplitude  with $J^z_{tot}= +{1}:   \gamma^*_T(S^z_\gamma = +1) + (S^z_{He}= 0 ) 
 \to  [(S^z_q  = + {1\over 2})  +  (S^z_{\bar q}  = - {1\over 2} )  + ( L^z_{q \bar q}= 0) ] _{\pi^0}   + (S^z_{He} =  0 )  + (L^z_{\pi^0 He^4} =1) $ 
vanishes at $t =0$, whereas the amplitude with a pion twist-3 amplitude 
$J^z_{tot}= +{1}:   \gamma^*_T(S^z_\gamma = +1) + (S^z_{He}= 0 ) 
 \to  [(S^z_q  = + {1\over 2})  +  (S^z_{\bar q}  = - {1\over 2} )  + ( L^z_{q \bar q}=+1) ] _{\pi^0}   + (S^z_{He} =  0 )  + (L^z_{\pi^0 He^4} =0  ) $ is finite at $t=0$,
thus discriminating between contributions using the twist-2 and twist-3 pion amplitudes.

\section{The Light-Front Vacuum}

It is important to distinguish the LF vacuum from the conventional instant-form vacuum.
The eigenstates of the instant-form Hamiltonian describe a state defined at a single instant of time $t$ over all space, and they are thus acausal as well as frame-dependent.  
The instant-form vacuum is defined as the lowest energy eigenstate of the instant-form Hamiltonian.
As discussed by Zee ~\cite{Zee:2008zz}, the cosmological constant  is of order $10^{120}$ times larger than what is observed if one computes the effects of quantum loops from QED.  Similarly, QCD instantons and condensates in the instant-form vacuum give a contribution of order  $10^{42}.$  The contribution of the  Higgs VEV computed in the instant form vacuum is  $10^{54}$ times too large. 

In contrast, the vacuum in LF Hamlitonian theory is defined as the eigenstate of $H_{LF}$ with lowest invariant mass.  It is defined at fixed LF time $\tau$ within the causal horizon, and  it is frame-independent; i.e., it is independent of the observer's motion.
Vacuum loop diagrams from quantum field theory do not appear  in the front-form vacuum since  the  $+$ momenta are positive: $k^+ _i = k^0_i+k^z_i\ge 0$, and the sum of $+$ momenta is conserved at every vertex.   The creation of particles cannot arise from the LF vacuum  since $ \sum_i  k^{+i}   \ne P^+_{vacuum} =0.$  
Since propagation with negative $k^+$  does not appear.  The physical vacuum state can also have $k^+=0$ modes corresponding to a flat energy-momentum background, analogous to a classical scalar
Stark or Zeeman field. For example,  Reinhardt and Weigl~\cite{Reinhardt:2012xs} have shown that the 
Nambu-Jona-Lasino   (NJL) model model can lead to  a nontrivial physical LF vacuum.
In the case of the Higgs theory, the traditional Higgs vacuum expectation value (VEV) is replaced by a ``zero mode"~\cite{Srivastava:2002mw}.  All phenomenological consequences of the Higgs theory in the Standard Model are unchanged in the LF formulation.  As noted in section 6, the $m^2\over x $ term in the LF kinetic energy $k^2_\perp + m^q\over x$ arises from the  interaction of a quark within a hadron in QCD with its Yukawa interaction with the Higgs background zero mode.

The physics associated with quark and gluon QCD vacuum condensates of the instant form are replaced by physical effects contained within the hadronic LFWFs. This is referred to as ``in-hadron" condensates~\cite{Casher:1974xd,Brodsky:2009zd,Brodsky:2010xf}.  For example, as discussed in section 7, the GMOR relation relating the vacuum-to-pion matrix elements of the axial current  and pseudoscalar operators is satisfied in LF theory as a relation between the twist-2 and twist-3 Fock states~\cite{Brodsky:2012ku}. 
The usual properties of chiral symmetry are satisfied, for example,  as discussed in section 2, the mass of the pion eigenstate computed from LF holography vanishes for zero quark mass.

The universe is observed within the causal horizon, not at a single instant of time.  The causal, frame-independent light-front vacuum can thus provide a viable match to the empty visible universe~\cite{Brodsky:2010xf}.    The huge contributions to the cosmological constant  from quantum field theory loops thus do not appear if one notes that the causal, frame-independent light-front vacuum has no quantum fluctuations --  in dramatic contrast to to  the acausal, frame-dependent instant-form vacuum;  the cosmological constant  arising from quantum field theory thus vanishes if one uses the front form.  The Higgs LF zero mode~\cite{Srivastava:2002mw}  has no energy-momentum density,  so it also gives zero contribution to the cosmological constant.  
The observed nonzero value could could be a property of gravity itself, such as the ``emergent gravity"  postulated by E. Verlinde~\cite{Verlinde:2016toy}. It is also possible that if one solves electroweak theory in a curved universe, the Higgs LF zero mode would be replaced with a field of nonzero curvature which could 
give a nonzero contribution  to the cosmological constant.

\section {The QCD Coupling at all Scales} 

The QCD running coupling $\alpha_s(Q^2)$
sets the strength of  the interactions of quarks and gluons as a function of the momentum transfer $Q$.
The dependence of the coupling
$Q^2$ is needed to describe hadronic interactions at 
both long and short distances. 
The QCD running coupling can be defined~\cite{Grunberg:1980ja} at all momentum scales from a perturbatively calculable observable, such as the coupling $\alpha^s_{g_1}(Q^2)$, which is defined from measurements of the Bjorken sum rule.   At high momentum transfer, such ``effective charges"  satisfy asymptotic freedom, obey the usual pQCD renormalization group equations, and can be related to each other without scale ambiguity by commensurate scale relations~\cite{Brodsky:1994eh}.  

\begin{figure}
\begin{center}
\includegraphics[height=10cm,width=15cm]{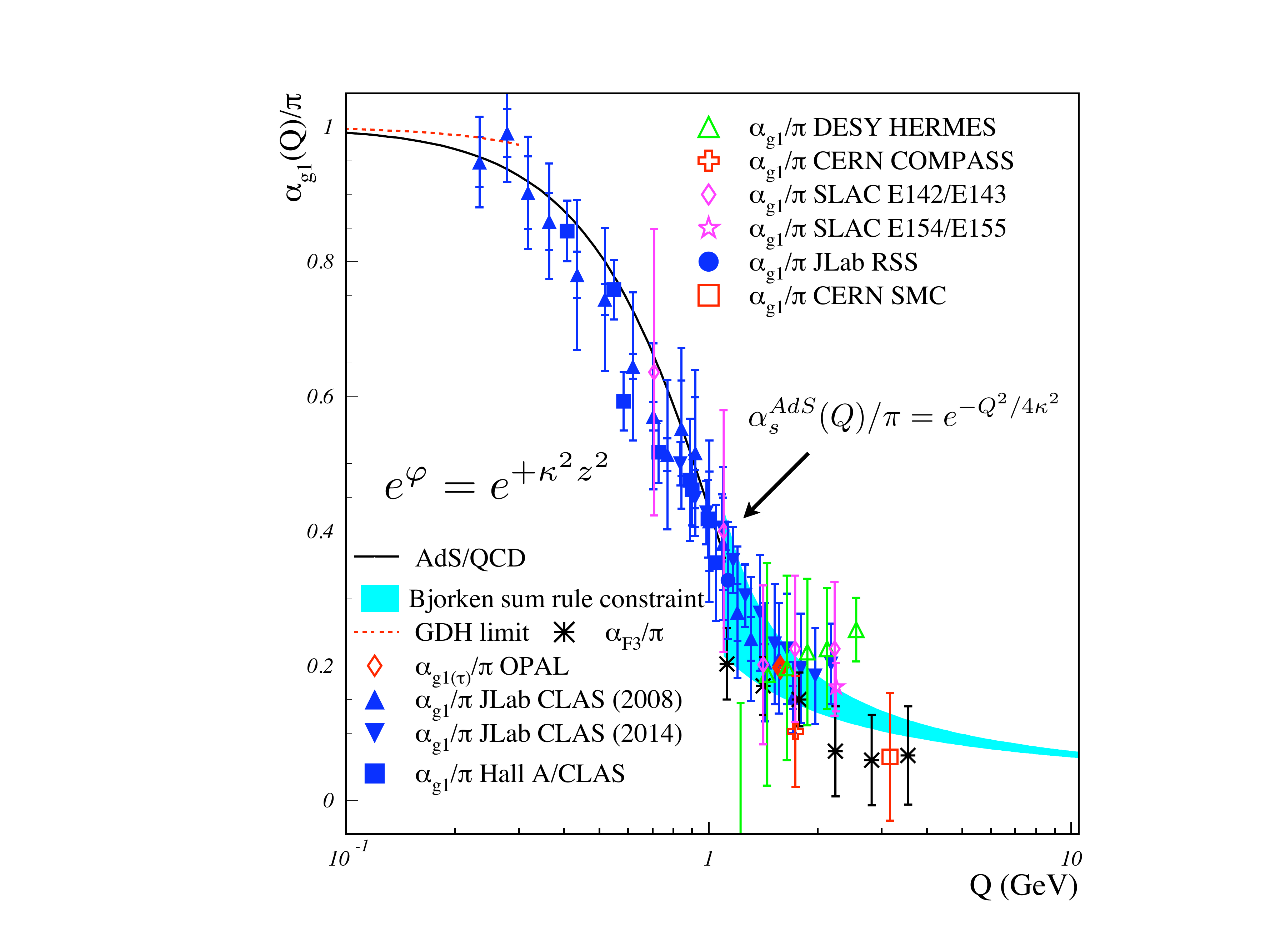}
\includegraphics[height=10cm,width=15cm]{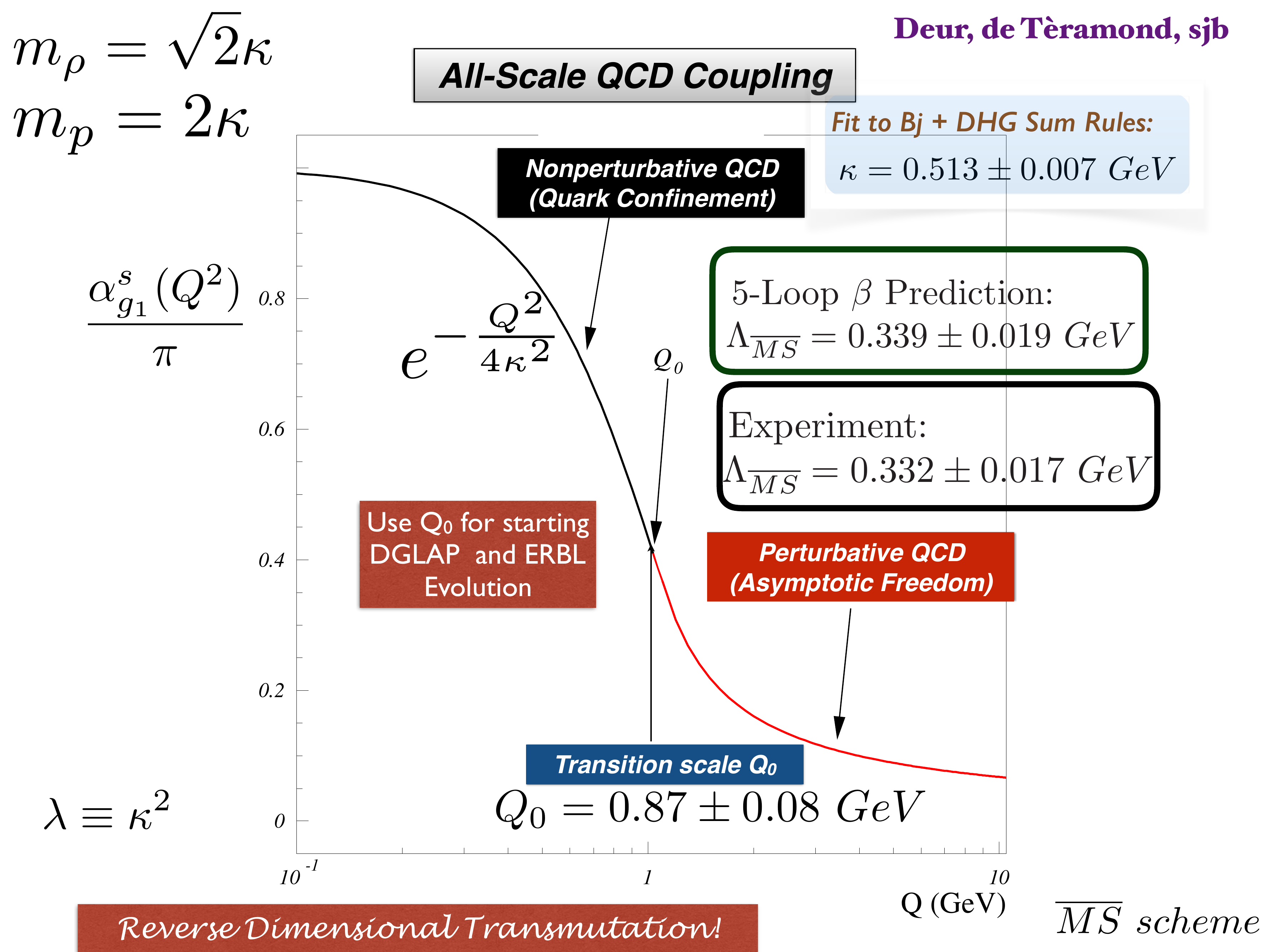}
\end{center}
\caption{
(A) Comparison of the predicted nonpertubative coupling, based on  the dilaton $\exp{(+\kappa^2 z^2)}$ modification of the AdS$_5$ metric, with measurements of the effective charge $\alpha^s_{g_1}(Q^2)$, 
as defined from the Bjorken sum rule.
(B)  Prediction from LF Holography and pQCD for the QCD running coupling $\alpha^s_{g_1}(Q^2)$ at all scales.   The magnitude and derivative of the perturbative and nonperturbative coupling are matched at the scale $Q_0$.  This matching connects the perturbative scale 
$\Lambda_{\overline{MS}}$ to the nonpertubative scale $\kappa$ which underlies the hadron mass scale. 
See Ref.~\cite{Brodsky:2014jia}. 
}
\label{DeurCoupling}
\end{figure}

The dilaton  $e^{+\kappa^2 z^2}$ soft-wall modification of the AdS$_5$ metric, together with LF holography, predicts the functional behavior of the running coupling
in the small $Q^2$ domain~\cite{Brodsky:2010ur}: 
${\alpha^s_{g_1}(Q^2) = 
\pi   e^{- Q^2 /4 \kappa^2 }}. $ 
Measurements of  $\alpha^s_{g_1}(Q^2)$ are remarkably consistent~\cite{Deur:2005cf}  with this predicted Gaussian form; the best fit gives $\kappa= 0.513 \pm 0.007~GeV$.   
See Fig.~\ref{DeurCoupling}
Deur, de Teramond, and I~\cite{Brodsky:2010ur,Deur:2014qfa,Brodsky:2014jia} have also shown how the parameter $\kappa$,  which   determines the mass scale of  hadrons and Regge slopes  in the zero quark mass limit, can be connected to the  mass scale $\Lambda_s$  controlling the evolution of the perturbative QCD coupling.  The high momentum transfer dependence  of the coupling $\alpha_{g1}(Q^2)$ is  predicted  by  pQCD.  The 
matching of the high and low momentum transfer regimes  of $\alpha_{g1}(Q^2)$ -- both its value and its slope -- then determines a scale $Q_0 =0.87 \pm 0.08$ GeV which sets the interface between perturbative and nonperturbative hadron dynamics.  This connection can be done for any choice of renormalization scheme, such as the $\overline{MS}$ scheme,
as seen in  Fig.~\ref{DeurCoupling}.  
The result of this perturbative/nonperturbative matching is an effective QCD coupling  defined at all momenta.   
The predicted value of $\Lambda_{\overline{MS}} = 0.339 \pm 0.019~GeV$ from this analysis agrees well the measured value~\cite{Agashe:2014kda}  
$\Lambda_{\overline{MS}} = 0.332 \pm 0.017~GeV.$
These results, combined with the AdS/QCD superconformal predictions for hadron spectroscopy, allow one to compute hadron masses in terms of $\Lambda_{\overline{MS}}$:
$m_p =  \sqrt 2 \kappa = 3.21~ \Lambda_{\overline{MS}},~ m_\rho = \kappa = 2.2 ~ \Lambda_{\overline{ MS} }, $ and $m_p = \sqrt 2 m_\rho, $ meeting a challenge proposed by Zee~\cite{Zee:2003mt}.
The value of $Q_0$ can be used to set the factorization scale for DGLAP evolution of hadronic structure functions and the ERBL evolution of distribution amplitudes.
Deur, de T\'eramond, and I have also computed the dependence of $Q_0$ on the choice of the  effective charge used to define the running coupling and the renormalization scheme used to compute its behavior in the perturbative regime.   
The use of  the scale $Q_0$  to  resolve  the factorization scale uncertainty in structure functions and fragmentation functions,  in combination with the scheme-indepedent {\it principle of maximum conformality} (PMC )~\cite{Mojaza:2012mf} for  setting   renormalization scales,  can 
greatly improve the precision of pQCD predictions for collider phenomenology.

\section{Is the Momentum Sum Rule Valid for Nuclear Structure Functions? }

Sum rules for deep inelastic scattering are usually analyzed using the operator product expansion of the forward virtual Compton amplitude, assuming it depends in the limit $Q^2 \to \infty$ on matrix elements of local operators such as the energy-momentum tensor.  The moments of structure functions and other distributions can then be evaluated as overlaps of the target hadron's light-front wavefunction,  as in the Drell-Yan-West formulae for hadronic form factors~\cite{Brodsky:1980zm,Liuti:2013cna,Mondal:2015uha,Lorce:2011dv}.
The real phase of the resulting DIS amplitude and its OPE matrix elements reflects the real phase of the stable target hadron's wavefunction.

The ``handbag" approximation to deeply virtual Compton scattering also defines the ``static"  contribution~\cite{Brodsky:2008xe,Brodsky:2009dv} to the measured parton distribution functions (PDF), transverse momentum distributions, etc.  The resulting momentum, spin and other sum rules reflect the properties of the hadron's light-front wavefunction.
However,  the final-state interactions which occur {\it after}  the lepton scatters on the quark, can give non-trivial contributions to deep inelastic scattering processes at leading twist and thus survive at high $Q^2$ and high $W^2 = (q+p)^2.$  For example, the pseudo-$T$-odd Sivers effect~\cite{Brodsky:2002cx} is directly sensitive to the rescattering of the struck quark. 
Similarly, diffractive deep inelastic scattering (DDIS)  involves the exchange of a gluon after the quark has been struck by the lepton~\cite{Brodsky:2002ue}.  In each case the corresponding DVCS amplitude is not given by the handbag diagram since interactions between the two currents are essential.
These ``lensing" corrections survive when both $W^2$ and $Q^2$ are large since the vector gluon couplings grow with energy.  Part of the final state phase can be associated with a Wilson line as an augmented LFWF~\cite{Brodsky:2010vs} which does not affect the moments.  

The Glauber propagation of the vector system $V$ produced by the  DDIS interaction on the nuclear front face and its subsequent  inelastic interaction with the nucleons in the nuclear interior $V + N_b \to X$ occurs after the lepton interacts with the struck quark.  The corresponding amplitude for deeply virtual Compton scattering is not given by the handbag diagram alone since interactions between the two currents are essential.
Because of the rescattering dynamics, the DDIS amplitude acquires a complex phase from Pomeron and Regge exchange;  thus final-state  rescattering corrections lead to  nontrivial ``dynamical" contributions to the measured PDFs; i.e., they are a consequence of the scattering process itself~\cite{Brodsky:2013oya}.  The $ I = 1$ Reggeon contribution to DDIS on the front-face nucleon then leads to flavor-dependent antishadowing~\cite{Brodsky:1989qz,Brodsky:2004qa}.  This could explain why the NuTeV charged current measurement $\mu A \to \nu X$ scattering does not appear to show antishadowing, in contrast to deep inelastic electron-nucleus scattering as discussed in ref. ~\cite{Schienbein:2007fs}.

Diffractive deep inelastic scattering is leading-twist, and it is an essential component of the two-step amplitude which causes shadowing and antishadowing of the nuclear PDF.  It is important to analyze whether the momentum and other sum rules derived from the OPE expansion in terms of local operators remain valid when these dynamical rescattering corrections to the nuclear PDF are included.   The OPE is derived assuming that the LF time separation between the virtual photons in the forward virtual Compton amplitude 
$\gamma^* A \to \gamma^* A$  scales as $1/Q^2$.
However, the propagation  =of the vector system $V$ produced by the DDIS interaction on the front face and its inelastic interaction with the nucleons in the nuclear interior $V + N_b \to X$ are characterized by a non-vanishing LF time interval in the nuclear rest frame.   
Note also  that shadowing in deep inelastic lepton scattering on a nucleus  involves  nucleons facing the incoming lepton beam. The  geometrical orientation of the shadowed nucleons  is not a property of the the  nuclear LFWFs used to evaluate the matrix elements of local currents.  Thus leading-twist shadowing and antishadowing appear to invalidate the sum rules for nuclear PDFs.  The same complications occur in the leading-twist analysis of deeply virtual Compton scattering $\gamma^* A \to \gamma^* A$ on a nuclear target. Thus the leading-twist multi-nucleon processes which produce shadowing and antishadowing in a nucleus are  not accounted for using the $Q^2 \to \infty$ OPE analysis.

\section{Summary} 

The light-front Hamiltonian equation $H_{LF}|\Psi> = M^2 |\Psi>$, derived from quantization at fixed LF time $\tau = t+z/c$ provides a causal, Poincar\'e--invariant, method for solving QCD. The eigenvalues $M^2_H$ are the squares of the hadronic masses, and the eigensolutions provide the LF Fock-state wavefunctions  $\Psi_n(x_i, \vec k_{\perp i}, \lambda_i) $  controlling hadron dynamics.   The  LFWFs $\Psi_n$ are independent  of the hadron's momentum; i.e.,  they are boost invariant and satisfy momentum and spin sum rules.  
Light-Front Quantization thus provides a physical, frame-independent formalism for hadron dynamics and structure.  Observables such as structure functions, transverse momentum distributions, and distribution amplitudes are defined from the hadronic light-front wavefunctions.

The full  QCD LF equation can be reduced for massless quarks to an  effective LF Shr\"odinger radial equation for the valence  $|q \bar q> $ Fock state of $q \bar q$ mesons
 $$[-{d^2\over d\zeta^2} + {4 L^2-1\over 4 \zeta^2 } + U(\zeta^2) ] \psi = M^2 \psi$$ 
 and similar bound-state equations for baryons, represented as  quark + diquark-cluster $ |q [qq]>$  eigenstates.
The  ``radial"  LF variable $\zeta^2= b^2_\perp x(1-x)$ of LF theory is conjugate  to the LF kinetic energy.  The identical equation is derived from $AdS_5$, where the fifth coordinate $z$ is identified with $\zeta$ (Light Front Holography). 

The color-confining potential $U(\zeta^2) = \kappa^4 \zeta^2 + 2\kappa^2(J-1) $ can be derived from soft-wall $AdS_5$ by incorporating the remarkable dAFF principle that a mass scale can appear in the Hamiltonian while retaining the conformal invariance of the action.  
The result is a color-confining LF potential which depends on a single universal constant $\kappa$  with mass dimensions.   In addition, by utilizing superconformal algebra~\cite{Dosch:2015bca}, the resulting hadronic color-singlet eigenstates have a $2 \times 2 $ representation of mass-degenerate bosons and fermions:   a $|q \bar q>$ meson with $L_M=  L_B+1$, a baryon doublet $|q [qq]>$ with 
$L_B$ and  $L_B+1$ components of equal weight,  and a tetraquark $|[qq] [\bar q \bar q] >$ with $ L_T = L_B$.  See: Fig. \ref{2X2Multiplets}.   Thus ratios of hadron masses such as $m_\rho = {M_p\over \sqrt 2}$ are predicted. The individual contributions  LF kinetic energy, potential energy, spin-interactions, and the quark mass to the mass squared of each hadron is also shown.  The virial theorem for harmonic oscillator confinement predicts the equality of the LF kinetic and potential contributions to $M^2_H$ for each hadron.

One obtains new insights into the hadronic 
spectrum, light-front wavefunctions, and the $e^{-{Q^2\over 4 \kappa^2}}$ Gaussian functional form of the QCD running coupling in the nonperturbative domain using light-front holography -- the duality between the front form   and AdS$_5$, the space of isometries of the conformal group. 
AdS/QCD also predicts the analytic  form of the nonperturbative running coupling  $\alpha_s(Q^2) \propto e^{-{Q^2\over 4 \kappa^2}}$,  in agreement with the effective charge  measured from measurements of the Bjorken sum rule.
This analysis also provides a connection between nonperturbative QCD  and PQCD at a scale $Q_0$ and  a prediction for $\Lambda_{\overline {MS}}$ from the proton or $\rho$ mass.

Other  LF Holographic predictions include:
\begin{enumerate}
\item Universal Regge-slopes in $n$ and $L$ for mesons:   $M^2(n,L) = 4\kappa^2(n+L)$ for mesons and $M^2(n,L) = 4\kappa^2(n+L+1)$ for baryons, consistent with measurements
\item The pion  eigenstate is  a massless $q \bar q$  bound state for chiral QCD $(m_q=0)$.
\item Empirically viable predictions for spacelike and timelike hadronic  form factors, structure functions, distribution amplitudes, and transverse momentum distributions~\cite{Sufian:2016hwn}
\item Superconformal extensions to heavy-light quark  mesons and baryons
\end{enumerate}

In addition, superconformal algebra leads to remarkable supersymmetric relations between mesons and baryons of the same parity.  The mass scale $\kappa$ underlying confinement and hadron masses  can be connected to the parameter   $\Lambda_{\overline {MS}}$ in the QCD running coupling by matching the nonperturbative dynamics, as described by  the effective conformal theory mapped to the light-front and its embedding in AdS space, to the perturbative QCD  regime. The result is an effective coupling  defined at all momenta.   This 
matching of the high and low momentum transfer regimes determines a scale $Q_0$ which  sets the interface between perturbative and nonperturbative hadron dynamics.  
The use of $Q_0$ to  resolve  the factorization scale uncertainty for structure functions and distribution amplitudes,  in combination with the principle of maximal conformality (PMC)  for  setting the  renormalization scales~\cite{Mojaza:2012mf},  can 
greatly improve the precision of perturbative QCD predictions for collider phenomenology.  The absence of vacuum excitations of the causal, frame-independent  front form vacuum has important consequences  for the cosmological constant.   I have also discussed evidence that the antishadowing of nuclear structure functions is non-universal; {\it i.e.},  flavor dependent, and why shadowing and antishadowing phenomena may be incompatible with sum rules for nuclear parton distribution functions.

Future work will include the extension of superconformal representations to pentaquark and other exotic hadrons, comparisons with lattice gauge theory predictions,  the construction of an  AdS/QCD orthonormal basis to diagonalize the QCD light-front hamiltonian,  hadronization at the amplitude level; and the computation of intrinsic heavy-quark higher Fock states.

\section*{Acknowledgments}

To appear in the proceedings of the 5th Winter Workshop on Non-Perturbative Quantum Field Theory (WWNPQFT) at the Universit\'e de Nice, Sophia Antipolis, France, March 22-24, 2017.   I thank Ralf Hofmann for organizing an outstanding meeting.
The results presented here are based on collaborations  and discussions  with  
Kelly Chiu, Alexandre Deur, Guy de T\'eramond, Guenter Dosch, Susan Gardner, Fred Goldhaber,  Paul Hoyer, Dae Sung Hwang,  
Rich Lebed, Simonetta Liuti, Cedric Lorc\'e, Valery Lyubovitskij,
Matin Mojaza, Michael Peskin, Craig Roberts, Robert Shrock, Ivan Schmidt, Peter Tandy, and Xing-Gang Wu.
This research was supported by the Department of Energy,  contract DE--AC02--76SF00515.  
SLAC-PUB-17012. 


\end{document}